\documentclass[sn-basic]{sn-jnl}
\jyear{2022}%

\theoremstyle{thmstyleone}%
\theoremstyle{thmstyletwo}%

\theoremstyle{thmstylethree}%

\begin{document}
	
	\title[Bayesian Optimal Fingerprinting]{Bayesian Quantification of Covariance Matrix Estimation Uncertainty in Optimal Fingerprinting}
	
	
	\author*[1]{\fnm{Samuel} \sur{Baugh}}\email{samuelbaugh@ucla.edu}
	
	\author[1,2]{\fnm{Karen A.} \sur{McKinnon}}\email{kmckinnon@ucla.edu}

	\affil*[1]{\orgdiv{Department of Statistics}, \orgname{University of California, Los Angeles}, \orgaddress{\street{ 520 Portola Plaza}, \city{Los Angeles}, \postcode{90095}, \state{Los Angeles}, \country{USA}}}
	
	\affil[2]{\orgdiv{Institute of the Environment and Sustainability}, \orgname{University of California, Los Angeles}, \orgaddress{\street{ 520 Portola Plaza}, \city{Los Angeles}, \postcode{90095}, \state{Los Angeles}, \country{USA}}}
	
	
	\abstract{Regression-based optimal fingerprinting techniques for climate change detection and attribution require the estimation of the forced signal as well as the internal variability covariance matrix in order to distinguish between their influences in the observational record. While previously developed approaches have taken into account the uncertainty linked to the estimation of the forced signal, there has been less focus on uncertainty in the covariance matrix describing natural variability, despite the fact that the specification of this covariance matrix is known to meaningfully impact the results. Here we propose a Bayesian optimal fingerprinting framework using a Laplacian basis function parameterization of the covariance matrix. This parameterization, unlike traditional methods based on principal components, does not require the basis vectors themselves to be estimated from climate model data, which allows for the uncertainty in estimating the covariance structure to be propagated to the optimal fingerprinting regression parameter. We show through a CMIP6 validation study that this proposed approach achieves better-calibrated coverage rates of the true regression parameter than principal component-based approaches. When applied to HadCRUT observational data, the proposed approach detects anthropogenic warming with higher confidence levels, and with lower variability over the choice of climate models, than principal-component-based approaches.}

	
	
	
	\maketitle
	
		\section{Introduction}\label{introduction_sec}
	
	While it is well-established that increasing greenhouse gas concentrations will lead to warming surface temperatures  \citep{arrhenius1896xxxi,masson2021ipcc}, it is difficult to quantify the extent to which changes throughout the observational record can be attributed to human influence. This issue is addressed through climate change detection and attribution (D\&A) research, where a combination of General Circulation Model (GCM) output and statistical techniques are used to separate the anthropogenic signal from natural variability in the observational record. ``Optimal fingerprinting'' is in particular concerned with identifying the relationship between global ``forcing'' patterns. Forcing is defined as any change which influences the radiative balance of the atmosphere, and while the primary forcing under consideration here is increasing greenhouse gas concentrations, other forcings include changing aerosol concentrations and changes in solar radiation due to natural cycles.  Optimal fingerprinting frameworks generally proceed by regressing the observed pattern of a particular climate variable onto a set of forcings estimated using climate models, with the residuals attributed to natural variability. Statistical inference on regression coefficients can then be used to assess the extent to which the forced signal can be ``detected'' in the observations and the degree to which observed trends can be ``attributed'' to the forcing. The regression framework has formed the conceptual foundation of optimal fingerprinting since it was introduced by \cite{hegerl1996detecting}, who used generalized least squares (GLS) to give the residuals a covariance structure that can be informed by GCM simulations of natural variability. 
	
	The reliability of optimal fingerprinting conclusions depends crucially on their ability to accurately represent uncertainty in the regression coefficient. This is complicated by the fact that various sources of uncertainty are present in the optimal fingerprinting procedure, including observational uncertainty, uncertainty in estimating the forced component, and uncertainty in representing the natural variability covariance matrix. In light of this, while the GLS framework takes into account the uncertainty induced by natural variability itself, there have been efforts in the optimal fingerprinting literature to expand the framework and incorporate additional sources of uncertainty. \citet{allen2003estimating} proposed the use of a total least squares inference approach to take into account the uncertainty induced by estimating the forced signal from a limited number of climate model runs. These approaches assume that the covariance structure is the same under forced and unforced climate scenarios. This assumption is relaxed in the methodology of \citet{huntingford2006incorporating} and \citet{hannart2014optimal} who allow for separate covariance matrices for different forcing patterns.

	A particular challenge with the regression approach is choosing how to represent the covariance matrix. The standard approach is to use a limited number of principal components estimated from the empirical covariance matrix of a set of climate model runs. The reason for truncating the principal components is that, while the first estimated principal components will generally correspond to meaningful patterns of variability, higher-number principal components may not correspond to meaningful patterns and their variances are likely to be under-estimated. The reason for this is that lower-variability directions of the sub-spaces are less likely to be represented within a finite number of control runs than higher-variability components, which leads to under-estimation of the lower-variance eigenvalues \citep{allen1999checking}. Under standard GLS regression formulas, terms are weighted according to the inverse of their variance, which is often called the ``precision''.  Under-estimated variances, or over-estimated precisions, will artificially increase the weight of non-meaningful patterns in the inference. A commonly used way to choose the number of principal components is the residual consistency test of \citet{allen1999checking}, which rejects components with variances that are unrealistically small as gauged by residuals from the regression model. Another approach to minimizing the effect of under-estimated variances is through regularization techniques. In particular, \citet{ribes2009adaptation} and \citet{Ribes2012} use the Ledoit and Wolf estimator \citep{LedoitWolf2004}. This method can be viewed as obtaining a full-rank covariance matrix from the empirical covariance matrix by scaling upward the variances of higher-number components while ``shrinking'' the variance of lower-number components under the constraint that the total variance remains the same.
	
	The approaches discussed above have used point estimators of the covariance structure that are estimated prior to model fitting and treated as fixed from the perspective of inference. This precludes the uncertainty from estimating the covariance structure from being represented in the uncertainty values for the final result. Recent developments have focused on developing detection and attribution methodology that performs covariance estimation and regression in an integrated fashion. \citet{hannart2016integrated} propose a frequentist integrated inference procedure using Ledoit-Wolf shrinkage of the covariance matrix. \citet{ribes2017new} develop an error-in-variables approach that is not based on the regression framework and is motivated by the incorporation of model uncertainty, meaning the uncertainty induced by the differences between climate models from each other and from reality, as opposed to just estimation uncertainty, which is a product of the limited number of climate model runs.
	
	The two integrated frameworks discussed above take a frequentist view of the optimal fingerprinting problem, where point estimators are developed based on observed data and confidence intervals are constructed based on their sampling distributions. Frequentist frameworks assume that the underlying quantity is fixed but unknown. However, the underlying quantities of the climate system could be more naturally thought of as random variables due to the underlying probabilistic nature of the climate system \citep{slingo2011uncertainty}. A Bayesian philosophy would be a natural way to integrate the various sources of information and uncertainty present into a statistical framework. This is the logic motivating \citet{katzfuss2017bayesian}, or KHS17 from here on out, who develop a hierarchical Bayesian approach for optimal fingerprinting.  This framework uses the truncated principal component parameterization of the covariance matrix where the component variances and the number of components are treated as variables. In the implementation, the component variances are estimated along with the regression parameters conditioned on each truncation number using MCMC. Then Bayesian model averaging (Bayesian model averaging; \citet{hoeting1999bayesian}) is used to marginalize over the number of components. This allows for the uncertainty in estimating these weights and the number of components in the covariance matrix to be propagated to the final inference results. A limitation to this method is that the principal components are computed in advance and treated as fixed in the inference, so only the uncertainty in estimating the variance coefficients is propagated to the final result. 
	
	The goal of integrated D\&A approaches is to correctly propagate the various sources of uncertainty to the final result. Better modeling of the uncertainty can lead to higher standard errors than approaches that omit sources of uncertainty or under-estimate component variances. The reliability of these standard errors can be gauged through coverage rates, defined as the percentage of the time that an uncertainty interval contains the true value when the true value is known. Ideally, a $90\%$ confidence interval will contain the true value $90\%$ of the time; if under-coverage is observed it indicates that the model is under-estimating variance or uncertainty. Fortunately, climate model ensembles provide a wealth of data for evaluating coverage rates of statistical methods under the assumption that the observations are truly generated from the appropriate distribution.
	
	Optimal fingerprinting methods have been observed to exhibit under-coverage under various settings. \citet{delsole2019confidence} show that confidence intervals are biased in a total least squares setting similar to those developed by \citep{allen2003estimating,hannart2014optimal}, and propose a bootstrapping method to correct for this bias. \citet{li2021uncertainty} demonstrate under-coverage when using the regularized method of \citep{Ribes2012}, and also propose to correct the standard errors using a bootstrapping approach. While these methods are able to achieve coverage rates closer to the nominal values, bootstrapping approaches produce uncertainty values that are no longer valid from the perspective of a probability model on the data. Under-coverage in the uncertainty intervals indicates that there are sources of variability that are not being taken into account, and an ideal approach would identify and statistically model these factors directly. Furthermore, these two papers are concerned with the frequentist properties of the detection and attribution regression coefficients, and such differ from the Bayesian approach considered here.

	This paper aims to develop optimal fingerprinting methodology within a Bayesian framework with improved coverage properties. The proposed approach builds off of the Bayesian framework of KHS17, with two key innovations. The first is the parameterization of the covariance matrix using Laplacian basis functions, rather than principal components, to avoid the uncertainty induced by estimating principal components from climate model data and to better propagate the uncertainty in estimating the covariance matrix to the final result. In addition, the model-driven component selection approaches using the normal regression likelihood model, such as the Bayesian model averaging approach of KHS17, are driven towards minimizing the residuals rather than accurately modeling the covariance matrix. This produces fits that achieve lower accuracy and misleading coverage when evaluated using climate models. To this end, the second proposed innovation is a separate Bayesian model for the number of components which uses a $\chi^2$ reparameterization of the likelihood to obtain an accurate representation of the variance terms. The overall proposed approach is to fit these two Bayesian models for the regression and the number of components simultaneously through an iterative process. It will be shown that this approach achieves higher accuracy and better-calibrated coverage rates than when the principal component parameterization or the normal distributional assumption for modeling the number of components is used.

	\section{Methods}\label{methods_sec}
	
	Let $\textbf{y}$ denote a vector of true climate observations. In general, this vector can be spatial or spatio-temporal, however here it will be assumed that this is a spatial vector giving values of a climate variable at each of $n_{ \text{grid}}$ grid-points on the sphere. In the spatial-only case, changes in the climate will be represented by taking each point of $\textbf{y}$ to be the linear regression coefficient of the climate variable of interest over time. The validation study in Section \ref{validation_study} and the application in Section \ref{application_sec} will be using trends in near-surface air temperatures, specifically the CMIP-standard variable ``tas'', over the 25-year period 1990-2015. However, optimal fingerprinting methodology can be applied to other climate variables such as precipitation \citep{lambert2003detection,zhang2007detection,min2011human,wan2015attributing}, zonally-averaged tropospheric temperatures \citep{santer2003contributions,santer2013identifying,santer2018human}, precipitable water \citep{ma2017detectable,zhang2019detection}, oceanic oxygen levels \citep{andrews2013detecting}, streamflow timing \citep{hidalgo2009detection}, vegetation changes \citep{mao2016human}, temperature extremes \citep{christidis2013role,kim2016attribution,lu2016anthropogenic,seong2021anthropogenic}, and various other climate indices \citep{timmermann1999detecting,roberts2012detectability,hobbs2015new,weller2016human}.
	
	In practice, the true vector $\textbf{y}$ will not be exactly known. Let $\{\textbf{y}_1,\dots,\textbf{y}_{n_{ \text{obs-ens}}}\}$ denote $n_{ \text{obs-ens}}$ vectors from an ensemble designed to represent observational uncertainty, for example from the $n_{ \text{obs-ens}}=200$ HadCRUT fields of near-surface air temperatures \citep{morice2021updated}. For the application in Section \ref{application_sec}, the observational ensemble mean $\frac{1}{n_{ \text{obs-ens}}}\sum_{k=1}^{n_{ \text{obs-ens}}}\textbf{y}_k$ will be used as a point-estimate of $\textbf{y}$. If desired, the Bayesian methods discussed can be augmented to include uncertainties obtained from the observational ensemble using the computationally efficient formulas of KHS17.
	
	Let $\textbf{x}$ denote the true field of climate trends under a  ``forcing scenario''. The forcing scenario here will exclusively refer to historical forcings, which include increasing levels of greenhouse gas concentrations but also aerosals and changes in solar radiation. However, this methodology would still hold if a different forcing scenario were to be used. It should be noted that most treatments of optimal fingerprinting allow for multiple forcing patterns to be included simultaneously, allowing for example greenhouse gas forcing and ``natural forcings'' to be treated separately. As natural forcings are less pronounced when trends are taken over 25-year periods, this work will treat the historical forcing patterns as representative of the effect of greenhouse gases, although extending the approach to include multiple distinct forcings would be straightforward. 

	Let $\mathcal{P}_c=\{\textbf{z}_1,\dots,\textbf{z}_{n_{\mathcal{P}_c}}\}$ and $\mathcal{H}_f=\{\textbf{x}_1,\dots,\textbf{x}_{n_{\mathcal{H}_f}}\}$ denote sets of $\mathbb{R}^{n_{ \text{grid}}}$ vectors of trends under ``control'' and ``forced'' simulations respectively, where $1\leq c\leq N_\mathcal{P}$  indexes over the $N_\mathcal{P}$ climate models used under pre-industrial forcing and $1\leq f\leq N_\mathcal{H}$ indexes over the $N_\mathcal{H}$ climate models used under historical forcings. The lower-case values $n_{\mathcal{P}_c}$ and $n_{\mathcal{H}_f}$ refer to the number of fields obtained from the $c$-th control and $f$-th sets of pre-industrial and historical vectors respectively. Intuitively, $\mathcal{P}$ stands for ``pre-industrial'', $\mathcal{H}$ stands for ``historical'', $c$ stands for ``control'', and $f$ stands for ``forced''. The dependence of $\textbf{x}$ and $\textbf{z}$ on $\mathcal{P}_c$ and $\mathcal{H}_f$ will be omitted when clear from context. It will be assumed that natural variability in the climate system has a multivariate normal distribution, or $\textbf{z}_i\sim N(0,C_{\mathcal{P}_c})$ for each $1\leq i\leq n_{\mathcal{P}_c}$ with covariance matrix $C_{\mathcal{P}_c}\in\mathbb{R}^{n_{ \text{grid}},n_{ \text{grid}}}$. Let $\textbf{x}_j\sim N(\textbf{x}_{\mathcal{H}_f},C_{\mathcal{P}_c})$ for each $1\leq j\leq n_{\mathcal{H}_f}$ where $\textbf{x}_{\mathcal{H}_f}$ is the ``true'' forced vector for the $f$-th historical climate model.  Note that the natural variability covariance matrix $C_{\mathcal{P}_c}$ is assumed to be unchanged by forcing; this assumption is often made in the literature, although approaches such as \citet{hannart2014optimal} relax this assumption. The dependence on the climate model used to generate $\mathcal{P}_c$ for covariance matrix $C$ will be dropped when clear from context. The climate models that will be used for generating $\mathcal{P}_c$ and $\mathcal{H}_f$ are displayed in Table \ref{climmod_summary}. These will be described in further detail in Section \ref{cmip6_data_sec}, but it will be noted here that the simulation data was restricted to cases where $n_{\mathcal{P}_c}>5$ and $n_{\mathcal{H}_f}>5$ to ensure that there are sufficient numbers of control and historical fields to be used in the procedure.

	\begin{table} 
		\centering
		\linespread{1}\small
		\begin{tabular}{rrr}
			\hline
			& Number piControl & Number historical \\ 
			\hline
			ACCESS-ESM1-5 & 36 & 40 \\ 
			AWI-CM-1-1-MR & 13 &  \\ 
			CanESM5 & 57 & 65 \\ 
			CESM2 & 43 &  \\ 
			CESM2-FV2 & 18 &  \\ 
			CESM2-WACCM & 17 &  \\ 
			CESM2-WACCM-FV2 & 19 &  \\ 
			FGOALS-g3 & 27 &  \\ 
			FIO-ESM-2-0 & 20 &  \\ 
			GISS-E2-1-G & 66 & 46 \\ 
			MIROC6 & 28 &  \\ 
			NESM3 & 16 &  \\ 
			NorCPM1 & 48 & 30 \\ 
			NorESM2-MM & 15 &  \\ 
			SAM0-UNICON & 27 &  \\ 
			UKESM1-0-LL & 61 &  \\ 
			GISS-E2-1-H &  & 25 \\ 
			MIROC-ES2L &  & 31 \\ 
			MPI-ESM1-2-LR &  & 29 \\ 
			\hline
		\end{tabular}
		\caption{Number of pre-industrial control and historical ensemble members used from CMIP6 climate models. Missing numbers indicate that either the scenario was not available for the given climate model or the number of useable 25-year trend fields was not greater than 5. }\label{climmod_summary}
	\end{table}

	With this set-up, the key formula driving optimal fingerprinting methodology is \begin{equation}
		\textbf{y}\sim \beta\textbf{x}+\epsilon \text{ where }\epsilon\sim N(0,C).\label{glseq}
	\end{equation} Here the parameter $\beta$ is interpreted as the ``detection and attribution'' parameter. Detection is generally taken to be a rejection of the hypothesis test $H_0:\beta>0$, and attribution is taken to be, conditional on detection, whether or not $\beta$ is statistically indistinguishable from one. In the latter case, $\beta>0$ indicates that the forcing pattern $\textbf{x}$ is meaningfully present in the observed response. When $\beta$ is statistically indistinguishable from one, it indicates that the discrepancy between the observations and the forced pattern is within the range of natural variability and that there is no evidence that other forcings are needed to explain observed trends.
	
	If $\textbf{x}$ and $C$ were known, \eqref{glseq} could be fit using (GLS), which is a well-understood methodology with closed-form frequentist estimators \citep{aitken1936iv} and was used by the original approach of \citet{hegerl1996detecting}. Crucially, however, the fact that these values are not known adds uncertainty to the inference which needs to be taken into account. Nevertheless, the following section will proceed with an overview of the frequentist properties of GLS. While this is different from the proposed Bayesian method where the covariance matrix is unknown, these properties will give intuition on the relationship between the covariance parameters and inference on the regression coefficient.

	\subsection{Generalized Least Squares}\label{gls_sec}
	
	Let $C\in\mathbb{R}^{n\times n}$ be a positive semi-definite matrix of rank $r$ for $r\leq  n$. Let $C=P\Lambda P^T$ denote the principal component or eigen-decomposition of $C$, where $P\in\mathbb{R}^{n\times r}$ denotes the column-matrix of principle component vectors and $\Lambda= \text{diag}(\lambda_1,\dots,\lambda_r)$ denotes the diagonal matrix of non-zero eigenvalues. Let $C^+\equiv P^T\Lambda^{-1}P$ denote the Moore-Penrose pseudo-inverse of $C$. Assuming that all components of $C$ and the vectors $\textbf{x}$ and $\textbf{y}$ are known, under the GLS formulation of Equation \eqref{glseq} the maximum likelihood estimator of $\beta$ is   \begin{equation}
			\hat{\beta}=\frac{\textbf{x}^TC^{+}\textbf{y}}{\textbf{x}^TC^{+}\textbf{x}}. \label{beta_mle}
		\end{equation}
	  It should be noted that when $C$ is full rank, or $r=n$, then $C^{-1}\equiv C^+$ can be substituted into the relevant equations.
	
	Denote the projection of $\textbf{x}$ and $\textbf{y}$ onto the eigenspace of $C$ as  
		\[\textbf{x}^*=P^T\textbf{x}  \text{ and }\textbf{y}^*=P^T\textbf{y}.\]
	  Then, the numerator of Equation \eqref{beta_mle} becomes $\textbf{x}^TC^{+}\textbf{y}=\textbf{x}^TP\Lambda^{-1}P^T\textbf{y}=(\Lambda^{-1/2}\textbf{x}^*)^T(\Lambda^{-1/2}\textbf{y}^*)$ and the denominator becomes $\textbf{x}^TC^{-1}\textbf{x}=\textbf{x}^TP\Lambda^{-1}P^T\textbf{x}=(\Lambda^{-1/2}\textbf{x}^*)^T(\Lambda^{-1/2}\textbf{x}^*)$. So, writing out these inner products as summations it can be seen that the MLE of $\beta$ is   \begin{equation}
			\hat{\beta}=\frac{\sum_{i=1}^{r}x_i^*y_i^*/\lambda_i}{\sum_{i=1}^{r}\left(x_i^*\right)^2/\lambda_i}\label{glsbeta}
	\end{equation}  and it can be seen that the standard error is  \begin{equation}
			se(\hat{\beta})=\left(\sum_{i=1}^{r}\left(x_i^*\right)^2/\lambda_i\right)^{-1/2}.\label{glsbetasd}
	\end{equation}  With these equations, the GLS formulation can be viewed as ordinary least squares with heteroskedastic errors performed on the patterns projected onto the eigenspace of $C$. In particular, Equation \eqref{glseq} can be rewritten as  \begin{equation}
			\textbf{y}^*\sim \beta \textbf{x}^*+\epsilon; \text{  }\forall_i\epsilon_i\sim N(0,\lambda_i). \label{glsproj}
	\end{equation}  When $C$ is rank deficient, then $\textbf{x}^*,\textbf{y}^*\in\mathbb{R}^r$, and the transformation implied by the basis vector matrix $P$ is a projection onto a lower-dimensional subspace. When $C$ is full rank, this projection can be seen as a rotation in a direction where the coordinates of the rotated vector are statistically independent.
	
	In practice, the covariance matrix $C$ is not known and must be estimated. Information about $C$ can be obtained from the empirical covariance matrix for the pre-industrial trend fields, defined as    \begin{equation}\hat{C}=\frac{1}{n_{\mathcal{P}_c}}\sum_{k=1}^{n_{\mathcal{P}_c}}(\textbf{z}_k-\overline{\textbf{z}})^T(\textbf{z}_k-\overline{\textbf{z}}).\label{empirical_covmat}
	\end{equation}  The rank of this matrix will be equal to the number of control vectors $n_{\mathcal{P}_c}$, which is generally much less than $n_{ \text{grid}}$; in the case considered here,  $n_{ \text{grid}}=2592$ and the largest value of $n_{\mathcal{P}_c}$ is 66 (Table \ref{climmod_summary}). Due to this rank deficiency, the empirical covariance matrix cannot be used directly, and a parameterization of the covariance matrix must be used instead.

	\subsection{Truncated principal Component Parameterization}\label{eof_basis_sect}
	
	The eigen-decomposition $C=P\Lambda P^T$ can be equivalently written as $C=\sum_{i=1}^r \lambda_ip_ip_i^T$, where $p_i$ denote the eigenvector columns of $P$. The idea behind the truncated eigenvector representation is to choose an integer $1\leq \kappa\leq n_{\mathcal{P}_c}$ and define the rank-$\kappa$ covariance matrix
	 \begin{equation} C_{\kappa; \text{PC}}\equiv\sum_{i=1}^\kappa\lambda_ip_ip_i^T\equiv P_\kappa\Lambda_\kappa P_\kappa^ T\label{c_kappa} 
	\end{equation}  where $P_\kappa$ is the column matrix of the first $\kappa$ principal components and $\Lambda_\kappa= \text{diag}(\lambda_1,\dots,\lambda_\kappa)$. Here $C$ is given the subscript ``PC'' to differentiate it from the Laplacian parameterization that will be introduced in the next section. This covariance matrix can then be used with Equation \eqref{beta_mle} and \eqref{glsbetasd} to obtain GLS results in the frequentist setting. Traditionally, the principle component vectors and the integer $\kappa$ are selected and treated as fixed prior to inference, and the uncertainty in their estimation is not propagated to the final conclusion. 
	
\begin{figure} 
		\centering
		\includegraphics[width=\linewidth]{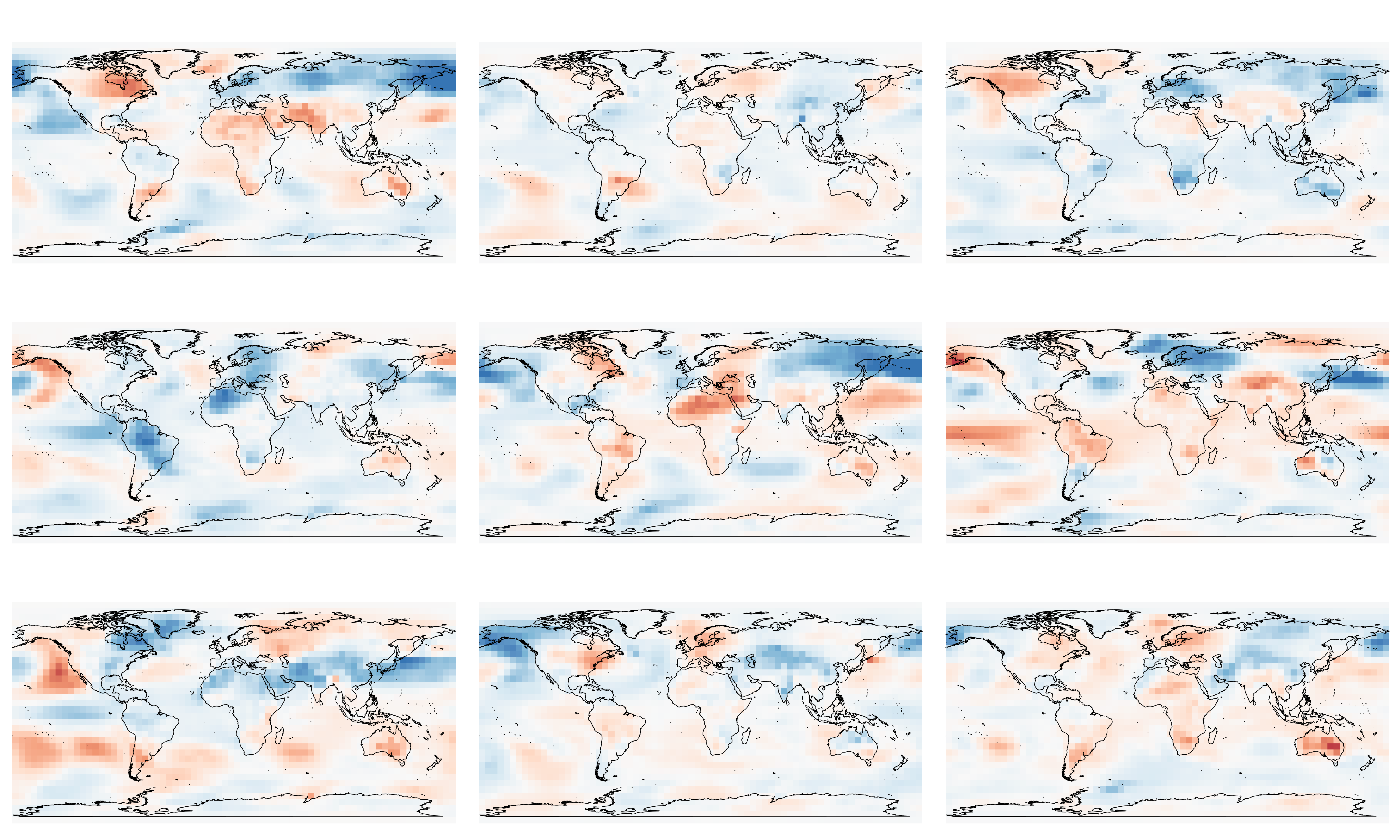}
		
		\includegraphics[width=.8\linewidth]{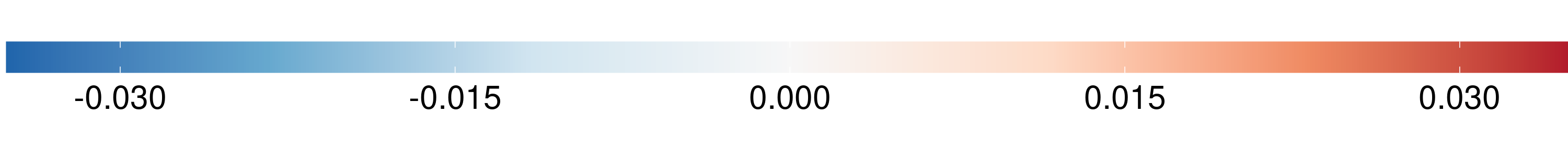} 
		\caption{Error in estimating the first principal component of the GISS-E2-1-G pre-industrial covariance matrix when only 13 control vectors are used. Displayed values are the difference between the principal components of nine random sub-samplings of the control vectors minus the first principal component when all 66 vectors are used. The number 13 was chosen to correspond to the minimum value of $n_{\mathcal{P}_c}$ in Table \ref{climmod_summary}. For comparison, the fully sampled principal component is displayed in Figure \ref{eoffigure}.} \label{eof_resamp}
	\end{figure}

\begin{figure} 
		\centering
		\includegraphics[width=\linewidth]{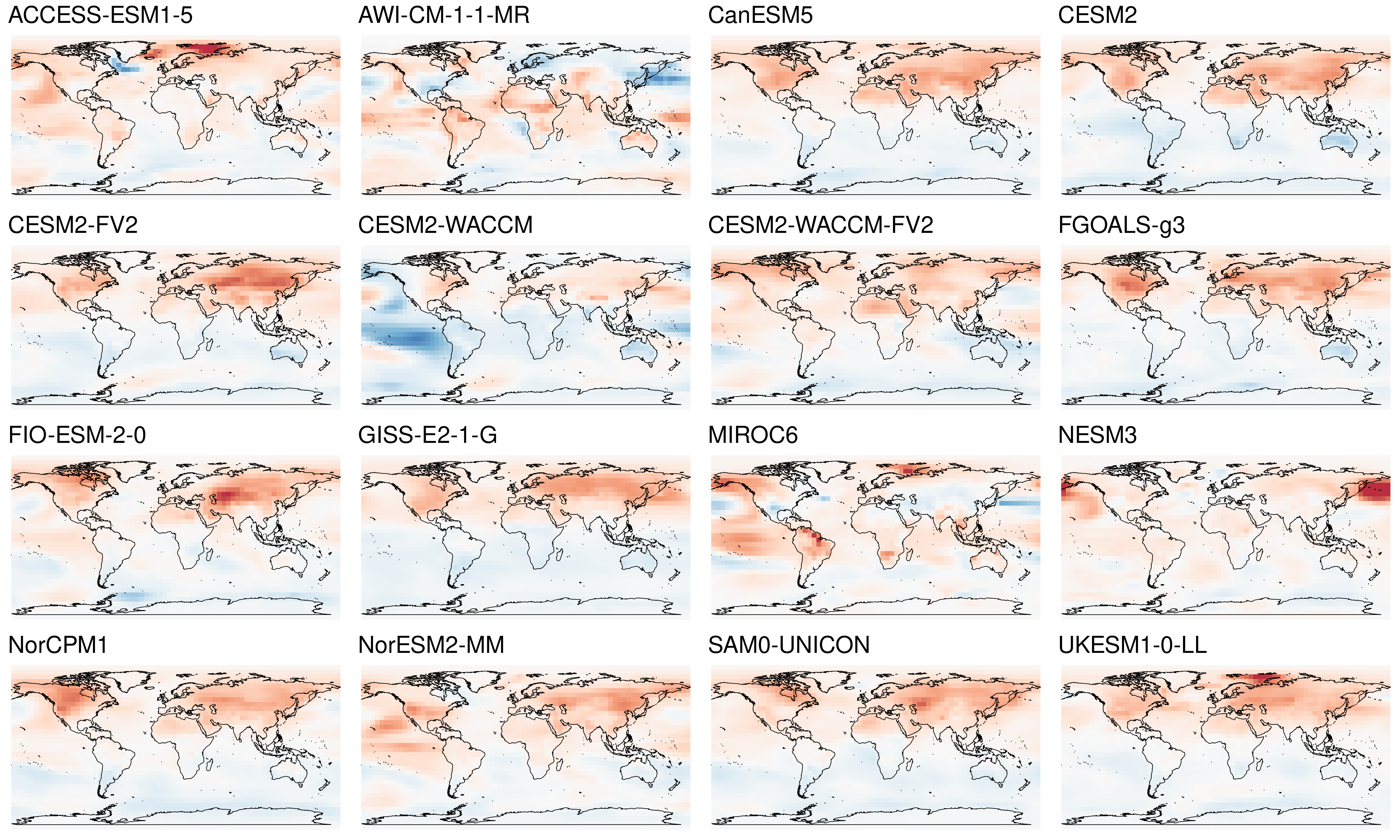}
		
		\includegraphics[width=.8\linewidth]{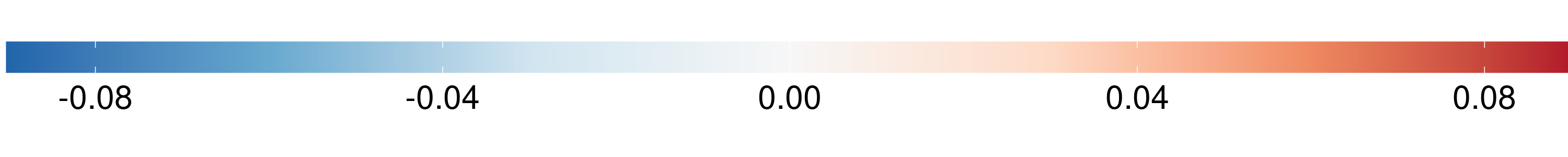} 
		\caption{First principal components, or EOFs, for the 16 pre-industrial climate ensembles used. Components were calculated using area-weighted fields of temperature trends over 1990-2015.} \label{eoffigure}
	\end{figure}

	One issue with treating the principal component vectors as fixed is that these components will contain estimation error due to the limited number of available control fields. To visualize this error, 13 trend fields were randomly sub-selected from the 66 trend fields of the GISS-E2-1-G model. Then the first principal component was computed from the sub-sampled vectors as well as the difference between this component and the first principal component obtained from using all 66 control vectors. This was repeated for nine random sub-samples with the results displayed in Figure \ref{eof_resamp}. These patterns can be interpreted as the error from estimating the first principal component using 13 control vectors if the covariance matrix implied by the full control vector set is treated as the ``truth''. The GISS-E2-1-G model was chosen as it has the highest number of pre-industrial control vectors, and the number 13 was chosen to correspond to AWI-CM-1-1-MR, which has the smallest number of control vectors (Table \ref{climmod_summary}). For reference, the first principal component for the ``true'' covariance of the GISS-E2-1-G model is displayed in the third row of Figure \ref{eoffigure}. It can be seen in Figure \ref{eof_resamp} that even for the first principal component there is notable error caused by the limited number of control runs. This visualization gives an idea of the scale and patterns of the error when estimating principal components from a limited number of control vectors; in reality, the true error for this number of components in this experiment would be larger due to the error between the covariance matrix implied by the 66 control vectors and the true covariance matrix.

	A second issue with the use of principal components is that they rely on the particular climate model used, and the covariance matrices for different climate models may differ from each other and from the ``true'' covariance structure. In particular, the ``true'' covariance matrix generating the field of observations can be imagined as being as different from the covariance matrices implied by the climate models as they are from each other. Under this conceptual framework, it can be seen that there may be a considerable ``mismatch'' between the components of the true, unobserved covariance matrix and those estimated using climate models, even when the number of principal components is restricted to the lower-number better-estimated patterns. Furthermore, in the inference procedure, these components are treated as fixed in advance, and there has not been any way developed to incorporate the uncertainty in their estimation into the optimal fingerprinting procedure. To get a sense of the difference between the covariance structures implied by the different climate models, Figure \ref{eoffigure} shows the estimated first principal components for each of the 16 models under consideration here (Table \ref{climmod_summary}). This shows notable differences between the patterns expressed in the first principal components of these different covariance matrices.

	The dependence of the principal component vectors on the size and properties of the climate models used presents a source of variability and uncertainty in the optimal fingerprinting procedure which is not easily represented. A third issue with the use of principal components is that they are not guaranteed to capture spatially coherent patterns. The first principal components visualized in Figure \ref{eoffigure} appear to capture spatially coherent coherent patterns, however this is a consequence of correlations present in the data as opposed to a property of the method. While it generally appears to be the case that lower-number principal components capture meaningful patterns, higher principal components are more likely to correspond to patterns that do not exhibit smooth spatial variation. Given the largely spatial nature of the fields under consideration in optimal fingerprinting, patterns that do not exhibit smooth spatial variation are likely to represent ``noise'' rather than meaningful modes of variability. This highlights the desirability of a basis function parameterization which enforces spatial correlation.
	
	\subsection{Laplacian Parameterization}\label{laplace_basis_sect}

	Given the issues with the principal component parameterization, it would be desirable to have a covariance matrix parameterization that relies on basis functions that do not depend on data. Ideally, these basis functions would be general enough to capture a wide potential variety of covariance patterns and would enforce spatial correlation in order to prioritize the inclusion of meaningful patterns. The eigenfunctions of the Laplace operator on the sphere yield basis vectors with these properties \citep{courant2008methods,delsole2015laplacian}. Specifically, the Laplace operator on the sphere is defined as  
		\[\nabla^2 Y=\frac{1}{\cos^2( \text{lat})}\left[\frac{\partial^2 Y}{\partial  \text{lon}^2}+\cos( \text{lat})\frac{\partial}{\partial  \text{lat}}\frac{\partial Y}{\partial  \text{lat}} \right]\]	  on latitudinal and longitudinal coordinates. Let $( \text{lat}_i, \text{lon}_i)$ denote the latitudinal and longitudinal coordinates of the $i$-th grid-cell for $1\leq i\leq n_{ \text{grid}}$. Assuming a grid with equally spaced latitudinal and longitudinal dimensions, let $\delta \text{lon}$ denote the distance between points in the longitudinal dimension and let $\delta \text{lat}$ denote the distance between points in the latitudinal dimension. Let $d_{gc}(i,j)$ denote the great circle distance between the $i$-th and $j$-th gridpoints. Then the discretized Laplacian operator can be represented by the matrix  $A\in\mathbb{R}^{n_{ \text{grid}}\times n_{ \text{grid}}}$ with elements   	\[ A_{ij}=\begin{cases} 
			-\frac{4}{\pi}\log(2(\sin(d_{gc}(i,j)^2)))\,\delta \text{lat}\,\delta \text{dlon}\,\sqrt{\cos( \text{lat}_i)\,\cos( \text{lat}_j)} &  \text{ for } i\neq j \\
			\frac{1}{4}\rho_i^2\left(1-2\log\frac{\rho_i}{\sqrt{2}}\right) & \text{ for } i=j
		\end{cases}
		\]	  where  \[\rho_i=\sqrt{\delta \text{lat}\,\delta \text{lon}\cos( \text{lat}_i)/\pi}\]   is the area of a circle whose area approximates the area of the $i$-th grid-cell.

	\begin{figure} 
		\centering
		\includegraphics[width=.95\linewidth]{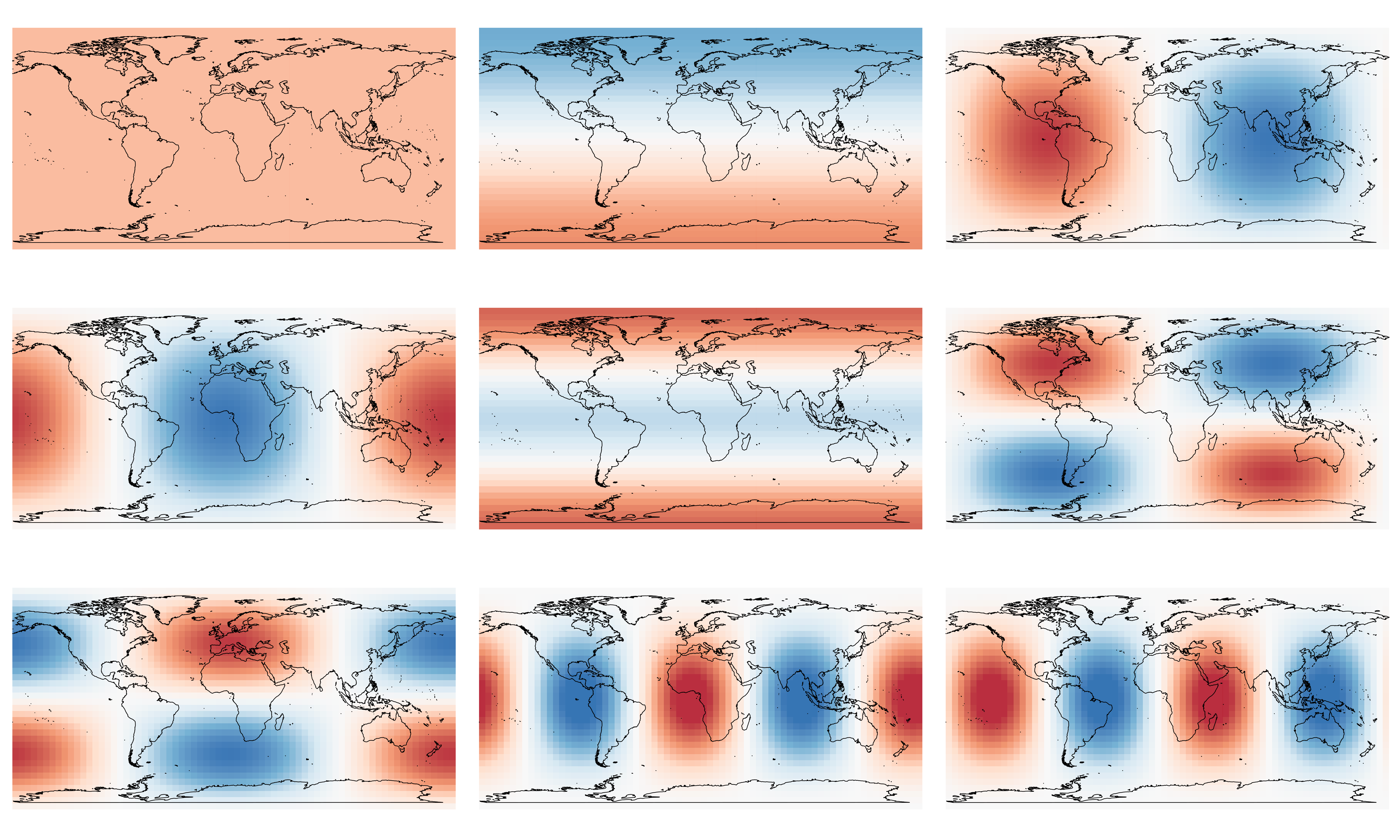}
		
		\includegraphics[width=.95\linewidth]{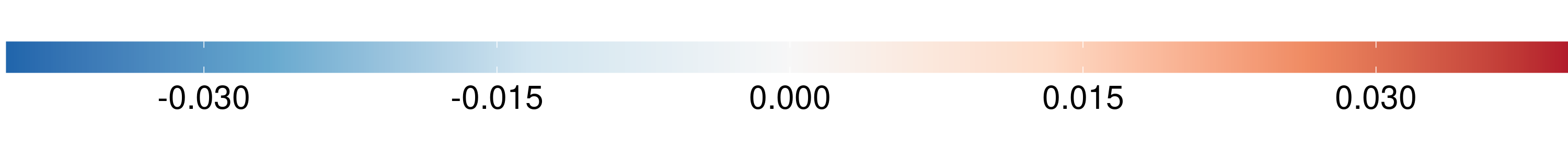} 
		\caption{First nine Laplacian basis functions on the sphere.}\label{lapfigure}
	\end{figure}
	
	Let $A^*$ denote $A$ projected onto the orthogonal component of the constant vector; this is done in order to ensure that the constant vector is an eigenfunction of the discretized Laplace operator. Let $\ell_1,\dots,\ell_{n_{ \text{grid}}}$ denote the normalized eigenvectors of the matrix $A^*$ where the constant pattern is re-ordered to come first. These vectors will be referred to as ``Laplacian basis vectors''. The reader is directed to \citet{delsole2015laplacian} for further details on their construction. To get a sense for the spatial relationships captured by these functions, the first nine eigenvectors on the sphere are displayed in Figure \ref{lapfigure}. Here it can be seen that these basis vectors capture spatially coherent correlation patterns at scales that decrease as the basis vector number increases.

	Let $L\in\mathbb{R}^{n_{ \text{grid}},n_{ \text{grid}}}$ denote the column vector matrix of $\ell_1,\dots,\ell_{n_{ \text{grid}}}$.  Define the Laplace-parameterized covariance matrix as $C_{ \text{Laplace}}=L\Lambda_{ \text{Laplace}} L^T=\sum_{i=1}^{n_{ \text{grid}}} \lambda_{i; \text{Laplace}}\ell_i\ell_i^T$ where $\Lambda_{ \text{Laplace}}= \text{diag}(\lambda_{1; \text{Laplace}},\dots,\lambda_{n_{ \text{grid}}; \text{Laplace}})$ is a diagonal matrix of variance coefficients for the Laplace basis vectors. The truncated Laplace covariance matrix is then defined as  \begin{equation}
			C_{\kappa; \text{Laplace}}\equiv\sum_{i=1}^\kappa\lambda_{i;{ \text{Laplace}}}\ell_i\ell_i^T\equiv L_\kappa\Lambda_{\kappa; { \text{Laplace}}} L^T_\kappa\label{lap_kappa}
	\end{equation}  where $\kappa$ is the truncated number of Laplace components, $L_\kappa$ is the column vector matrix of the first $\kappa$ Laplace eigenfunctions, and $\Lambda_{\kappa;{ \text{Laplace}}}= \text{diag}(\lambda_{1;{ \text{Laplace}}},\dots,\lambda_{\kappa;{ \text{Laplace}}})$. 
	
	As before, let $\{\textbf{z}_1,\dots,\textbf{z}_{n_{\mathcal{P}_c}}\}$ denote a collection of control-run vectors for pre-industrial climate simulation $\mathcal{P}_c$. Let $\textbf{z}_k^*=L^T\textbf{z}$. Then the empirical Lapalacian variance coefficients are  \begin{equation}
			\hat{\Lambda}_{ \text{Laplace}}= \text{diag}\left(\frac{1}{n_{\mathcal{P}_c}}\sum_{k=1}^{n_{\mathcal{P}_c}}\left(\textbf{z}_k^*-\overline{\textbf{z}^*}\right)^T\left(\textbf{z}_k^*-\overline{\textbf{z}^*}\right)\right) \text{ where }\textbf{z}_k^*=L^T\textbf{z}	\label{empirical_laplace_coefs}\end{equation}  and $\overline{\textbf{z}^*}=\frac{1}{n_{\mathcal{P}_c}}\sum_{k=1}^{n_{\mathcal{P}_c}}\textbf{z}_k^*$. The diagonal entries of $\hat{\Lambda}$ will be denoted as $\hat{\Lambda}_{ \text{Laplace}}=(\hat{\lambda}_{1; \text{Laplace}},\dots,\hat{\lambda}_{n_{ \text{grid}}; \text{Laplace}})$. This construction is analagous to the principal component case if $L$ is replaced with $P$. To see this, left-multiplying Equation \eqref{empirical_covmat} by $P^T$ and then right-multiplying the result by $P$ yields $P^T\hat{C}P=\Lambda=\left(\frac{1}{n_{\mathcal{P}_c}}\sum_{k=1}^{n_{\mathcal{P}_c}}\left(\textbf{z}_k^*-\overline{\textbf{z}^*}\right)^T\left(\textbf{z}_k^*-\overline{\textbf{z}^*}\right)\right)$ which is a diagonal matrix by construction.
	
	\begin{figure} 
		\centering
		\includegraphics[width=.9\linewidth]{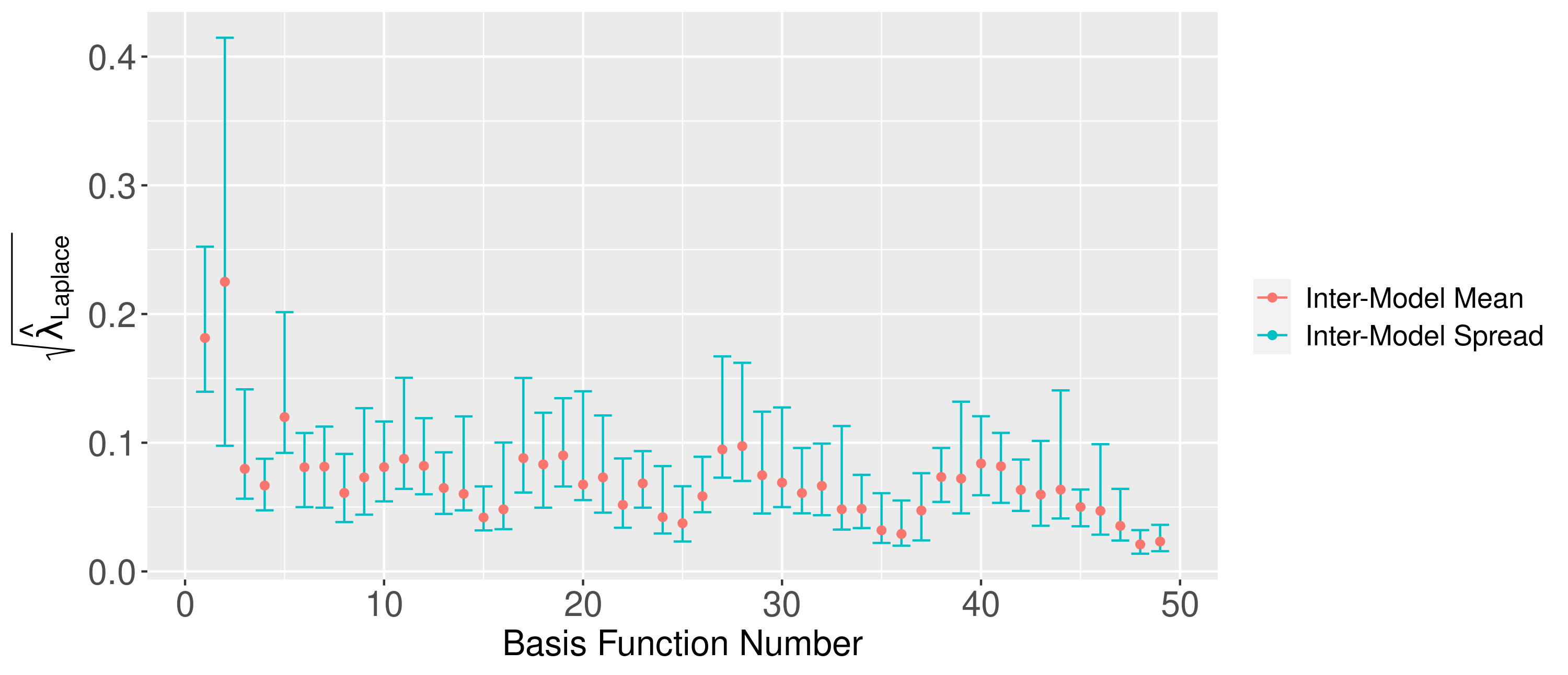}
		\label{cov_spread}
		\caption{Empirical standard deviations for the first fifty Laplacian basis functions. Inter-model means for each coefficient are over the 16 pre-industrial control models of Table \ref{climmod_summary}. Intervals show the maximum and minimum values of the coefficients over the 16 models.  }\label{lapcoefs}
	\end{figure}

	To get a sense for the values of the Laplacian variance coefficients, Figure \ref{lapcoefs} displays the values of $\sqrt{\hat{\lambda}_i}$, which represent the standard deviation for the $i$-th Laplacian component, for $1\leq i\leq 50$.  For each component, means and max-min spreads calculated over the $16$ pre-industrial control models are displayed. It can be seen that there remains a noticeable degree of inter-model heterogeneity; although unlike with the principal components displayed in Figure \ref{eoffigure} these differences can be visualized succinctly due to the fact that the underlying basis function representations are the same between models. There also appears to be a periodic oscillation in the Laplacian coefficients. This is due to the fact that the Laplacian basis vectors oscillate between primarily longitudinal correlation patterns and primarily latitudinal correlations as can be seen in Figure \ref{lapfigure}. The former correlation patterns are more prominent in the data since temperature changes are more correlated across longitude than across latitude.

	\subsection{Bayesian Hierarchical Model}\label{bayesian_sec}
	
	In the previous two sections $P$ has denoted the column matrix of principal component basis vectors and $L$ has denoted the matrix of Laplacian basis vectors. For the remainder of this section $B$ will denote a generic set of basis functions standing for either principal components or Laplacian basis vectors depending on which method is used. Let $\mathcal{K}$ denote a random variable for the number of components with discrete probability distribution $P(\mathcal{K}=\kappa)$ over $1\leq\kappa \leq n_{ \text{basis}}$; here  $n_{ \text{basis}}= \text{min}(n_{\mathcal{P}_c},n_{ \text{grid}})$ in the principal component case and $n_{ \text{basis}}=n_{ \text{grid}}$ in the Laplacian case. Conditioning on $\mathcal{K}=\kappa$, let $C_\kappa=B^T_\kappa\Lambda_\kappa B_\kappa$ represent the truncated covariance matrix with generic basis vectors and coefficients $\Lambda= \text{diag}(\lambda_1,\dots,\lambda_{\kappa})$. Let $\hat{\lambda}_1,\dots,\hat{\lambda}_{n_{ \text{grid}}}$ refer to either the eigenvalues of the empirical covariance matrix $\hat{C}$ defined in Equation \eqref{empirical_covmat} when the principal component basis is used or the empirical Laplacian coefficients defined in Equation \eqref{empirical_laplace_coefs} when the Laplacian basis is used. As in Section \ref{gls_sec}, let $\textbf{y}^*=B^T\textbf{y}$ and let $\textbf{x}^*=B^T\textbf{x}$.

	Conditioned on $\mathcal{K}=\kappa$, the Bayesian optimal fingerpriting model can be defined concisely with the following equations:  \begin{equation}
			{y}^*_i-\beta {x}^*_i\sim N(0,\lambda_i) \text{ for }1<i\leq \kappa \label{normeq}
	\end{equation}  \[\log\lambda_i\sim N(\log\hat{\lambda}_i,1) \text{ for }1<i\leq \kappa\]  \[\beta\sim  \text{Unif}(- \text{Inf}, \text{Inf})\]  where regression parameter $\beta$ is given a dispersed, uninformative prior in order to avoid biasing the results towards a particular value. 
	
	This basic framework is similar to and inspired by the approach of KHS17. One difference is that the above ignores the fact that \textbf{y} and \textbf{x} come from ensembles and rather takes them to be fixed at the respective ensemble means.  This simplification with regards to the forced ensemble would not be expected to affect the results, since the approach of KHS17 integrates the forced variability out of the likelihood function and as such it does not influence the posterior distributions. Observational uncertainty is likely to be small in the field of 25-year trends over the time period under consideration. As such this uncertainty is omitted here, although in principle the likelihood techniques of KHS17 could be used to represent observational uncertainty in a computationally efficient way. Another difference between our proposed framework and that of KHS17 is that their approach introduced a parameter $\sigma^2>0$ to be added to the diagonal of the covariance matrix in order to ensure a full-rank representation, whereas our approach uses a rank-deficient covariance structure.

	While using a rank-deficient covariance matrix in the frequentist context yields the closed-form estimators presented in Section \ref{gls_sec}, there is a slight complication from a Bayesian perspective as the likelihood for $y_i-\beta x_i$ in Equation \eqref{normeq} is not defined if $i>\kappa$. A way to get around this is to let $\textbf{1}\{i\leq \kappa\}$ denote the indicator function for $i\leq \kappa$ and write Equation \eqref{normeq} as  \begin{equation}
			({y}^*_i-\textbf{1}\{i\leq \kappa\}\beta_{\kappa} {x}^*_i)\sim N(0,\lambda_{i;\kappa}) \text{ for }1\leq i\leq n_{ \text{grid}} \label{indeq}
	\end{equation} 
	where $\kappa$ is added as a subscript to $\beta$ and $\lambda_i$ in order to represent the dependency between these parameters.

	\subsection{Selecting Covariance Truncation Number}\label{compsel_sec}

		Selection of the number of basis functions in the covariance matrix representation has long been identified as an important issue which significantly impacts the results of optimal fingerprinting \citep{hegerl1996detecting}. Within the frequentist framework the most common method employed for selecting the number of components is the ``residual consistency'' test originally proposed by \citep{allen1999checking}. This test is also used in the Bayesian framework of KHS17 for post-hoc model validation. The motivation for the test relies on the observation that when the covariance structure is being modeled accurately, the squared scaled and projected residuals will be $\chi^2$ distributed. To see this, recall that $\textbf{y}^*=B^T\textbf{y}$ and $\textbf{x}^*=B^T\textbf{x}$ are the observed and forced vectors respectively projected onto the basis functions implied by the covariance matrix parameterization. Let $r_{\kappa;i}=\frac{y_i^*-\beta x_i^*}{\sqrt{\lambda_i}}$ denote the normalized residual conditioned on a value of $\beta$. It can be seen from Equation \eqref{normeq} that  \begin{equation*}
			r_{\kappa;i}\equiv \frac{y_i^*-\beta x_i^*}{\sqrt{\lambda_i}}\sim N(0,1).
	\end{equation*}  Define the vector of normalized residuals as $\textbf{r}_{\kappa}=[r_{\kappa;1},\dots,r_{\kappa;\kappa}]=\Lambda^{-1/2}_\kappa B_\kappa^T(\textbf{y}-\beta \textbf{x})$. It follows that, conditioned on $\beta$, $(r_{\kappa;i})^2\sim\chi^2_1$, and that
	
	 \begin{equation}
			\textbf{r}_{\kappa}^T\textbf{r}_{\kappa}=\sum_{i=1}^\kappa (r_{\kappa;i})^2=\sum_{i=1}^\kappa\frac{(y_i^*-\beta x_i^*)^2}{\lambda_i}\sim \chi^2_k \label{chi2eq}
	\end{equation}    When not conditioning on $\beta$, the degrees of freedom are reduced by one, yielding $\textbf{r}_{\kappa}^T\textbf{r}_{\kappa}=(\textbf{y}-\beta \textbf{x})C^{-1}_\kappa(\textbf{y}-\beta \textbf{x})\sim \chi^2_{\kappa-1}$ which is the form of the test stated in Equation 18 of \citet{allen1999checking} when the number of forcing patterns is equal to one.

	The residual consistency test is then rejected for a particular $\kappa$ if $\textbf{r}_{\kappa}^T\textbf{r}_{\kappa}$ exceeds an $\alpha$-significance threshold with respect to the $\chi^2$  distribution. Looking at the summands of Equation \eqref{chi2eq}, it can be seen that large values will be achieved when the component variances $\lambda_i$ are small relative to the squared residuals $(y_i^*-\beta x_i^*)^2$. As such, this test is more likely to reject the number of components if the values of $\lambda_i$ under-estimate the true variance. The motivation for the test rejecting only high values of the test statistic is to avoid misleading and over-confident conclusions, however in principle a two-sided test could be used to ensure that the component variances are not over-estimated as well. The number of components is generally obtained by evaluating the results of the test for increasing values of $\kappa$ until the test is rejected, at which point $\kappa$ is fixed at the previous passing value.
	
	While useful, the residual consistency test carries significant drawbacks. One drawback is that the significance level $\alpha$ is somewhat arbitrary. A more conceptual difficulty is that in the hypothesis testing framework there is a difference between failing to reject a test and the null hypothesis being true. In the case of the residual consistency test, the null hypothesis is that the squared scaled and de-correlated residuals are $\chi^2$ distributed, which can be viewed as saying that the covariance matrix structure is accurate. A rejection of this test means that there is sufficient evidence from the observations to show that the estimated covariance structure is inadequate. Failure to reject the test means that there is not sufficient evidence to reject, but does not mean that the covariance structure is itself adequate. A third drawback of this procedure is that it is generally done prior to and separately from inference, and as such the uncertainty in choosing the number of components in this way is not incorporated into the model. Finally, concern has been raised that the statistical assumptions underlying the validity of the test in a frequentist framework may not be satisfied in practice \citep{mckitrick2022checking}. Despite these drawbacks, the residual consistency test remains a well-motivated approach to choosing the number of components to minimize underestimation of the component variances.

		A Bayesian approach that incorporates component selection within the inference procedure has the potential to remedy these drawbacks. This is the motivation for the framework proposed by KHS17, who estimate the distribution over the number of components by fitting the regression model separately for each value of $\mathcal{K}=\kappa$.  With these fitted models, Bayesian model averaging \citep{hoeting1999bayesian} is used to marginalize over $\mathcal{K}$ and obtain a posterior distribution for $\beta$ that takes the uncertainty in $\mathcal{K}$ into account. The primary issue with the Bayesian model averaging approach to selecting $\kappa$ is that its credible intervals tend to yield sub-nominal coverage rates when evaluated using climate model historical runs as surrogate observations, which will be discussed further in Section \ref{results_sec}. It appears that the normal likelihood term leads to ``over-fitting'' the data and selecting more than an optimal number of components. This results in suboptimal estimation of $\beta$ due to the fact that while higher-number components may increase the strength of the fit from the perspective of the likelihood model, if these components correspond to non-meaningful ``noise'' vectors their inclusion will lead to poor accuracy at recovering the true parameter. At the same time, increasing the number of components increases the model's confidence in its estimation of $\beta$, which results in sub-nominal coverage rates. These observations motivate the proposed model for $\mathcal{K}$ in the following section.

\subsection{Bayesian Model for Component Selection}\label{procedure_sec}

The Bayesian model averaging procedure for obtaining the posterior distribution of $\mathcal{K}$ uses weights which prioritize better fits of the regression likelihood term $({y}_i^*-\beta {x}_i^*)\sim N(0,\lambda_i)$. As discussed in Section \ref{gls_sec}, accurate estimates of the covariance components are essential to producing reliable estimates and uncertainty intervals. To obtain a distribution over $\mathcal{K}$ that models the fit of the covariance parameters we propose a reparameterization of the data distributional assumption of $\forall_{i=1}^\kappa(\textbf{y}_i^*-\beta \textbf{x}_i^*)\sim N(0,\lambda_i)$ to $\sum_{i=1}^\kappa \frac{(\textbf{y}_i^*-\beta \textbf{x}_i^*)^2}{\lambda_i}\sim \chi^2_\kappa$. This yields the following conditional likelihood model for $\mathcal{K}$:  \begin{equation}
		P(\mathcal{K}=\kappa\vert \textbf{y},\textbf{x},\theta)\propto [\textbf{y},\textbf{x}\vert \mathcal{K}=\kappa,\theta]=\chi^2\left(\textbf{r}_{\kappa}^T\textbf{r}_\kappa\vert df=\kappa-1 \right) \label{kappa_chi2}
\end{equation}  where the distribution of $\textbf{r}_\kappa=\Lambda^{-1/2}_\kappa B_\kappa^T(\textbf{y}-\beta \textbf{x})$  is obtained from Equation \eqref{chi2eq}; note here that the dependence of $\textbf{r}_\kappa$ on the parameter vector $\theta=(\beta,\lambda_1,\dots,\lambda_\kappa)$ is rendered implicit. This re-parameterization is inspired by the residual consistency test of \citet{allen1999checking} which yields higher likelihoods when then squared residuals and the component variances are of similar magnitude.

While this reparameterization is motivated by the need to accurately model $\mathcal{K}$, it would be misleading to use this likelihood model for $\beta$ itself since the mode of \eqref{kappa_chi2} over $\beta$ would no longer have a least-squares interpretation. As such, the proposed method of inference is to fit two inter-dependent Bayesian models, one for $\mathcal{K}$ and the other for $\theta$, each of which is conditioned on the value of the other. Under this separation, there becomes two choices for the distribution over $\mathcal{K}$, namely the normal regression parameterization and the $\chi^2$ reparameterization. The former choice yields the following posterior distribution for $\mathcal{K}$ when conditioned on $\theta$:  \begin{equation}
		P(\mathcal{K}=\kappa\vert \textbf{y},\textbf{x},\theta)\propto \prod_{i=1}^{n_{ \text{grid}}}\mathcal{N}({y}_i^*\vert \textbf{1}\{i\leq\kappa\}\beta {x}_i^*,\lambda_i). \label{kappa_norm}
\end{equation}   The latter choice yields the posterior distribution in Equation \eqref{kappa_chi2}. In the following, the first option of Equation \eqref{kappa_chi2} will be denoted as $\chi^2(\textbf{r}_\kappa^T\textbf{r}_\kappa\vert df=\kappa)\equiv \ell_{\chi^2}(\kappa,\theta)$ and the second option of Equation \eqref{kappa_norm} will be denoted as $\prod_{i=1}^{n_{ \text{grid}}}\mathcal{N}({y}_i^*\vert \textbf{1}\{i\leq\kappa\}{x}_i^*,\lambda_i)\equiv \ell_\mathcal{N}(\kappa,\theta)$. where the dependence on $\textbf{y}$ and $\textbf{x}$ is made implicit. These two choices will be generically denoted as elements of the set $\ell\in\{\ell_{\chi^2}$,$\ell_\mathcal{N}\}$. The $\ell_\mathcal{N}$ case can be viewed as equivalent to the Bayesian model averaging procedure if a flat prior was to be used for each $\lambda_i$.

In summary, the proposed method involves two Bayesian models each conditioned on the other; the first is the regression model for $\theta=(\beta,\lambda_1,\dots,\lambda_\kappa)$ conditioned on $\mathcal{K}=\kappa$ defined by Equation \eqref{indeq} and the second is the model for $\mathcal{K}$ conditioned on $\theta$ defined by either Equation \eqref{kappa_norm} in the $\ell_\mathcal{N}$ case or by Equation \eqref{kappa_chi2} in the $\ell_{\chi^2}$ case. The posterior distributions for each of these models can be evaluated simultaneously through the following iterative procedure. As input, this procedure takes the observational field $\textbf{y}$, the forced field $\textbf{x}$, a set of basis vector $B$ taken to be either principal components or Laplacian eigenfunctions, empirical variance terms $\hat{\lambda}_1,\dots,\hat{\lambda}_{n_{ \text{basis}}}$ calculated from control vectors, a choice of likelihood function for $\mathcal{K}$  out of the two choices $\ell\in\{\ell_\mathcal{N},\ell_{\chi^2}\}$, a function $g$ which returns a value from a distribution, and a number of MCMC samples $M$. The method proceeds as follows:

\begin{enumerate}
	\item Fix a choice of likelihood function $\ell$, which generically denotes either $\ell\equiv\ell_\mathcal{N}$ or $\ell\equiv\ell_{\chi^2}$. Let $\theta^{(0)}=(\beta^{(0)},\hat{\lambda}_1,\dots,\hat{\lambda}_{n_{ \text{basis}}})$ denote the ``first guess'' values for the regression parameters.  Using these values, evaluate the distribution $P(\mathcal{K}=k\vert \textbf{y},\textbf{x},\theta^{(0)})\propto \ell(\kappa,\theta^{(0)})$ for each $1<\kappa\leq n_{ \text{basis}}$; note that the restriction $1<\kappa$ is imposed to avoid degeneracy in the estimation of $\beta$.
	\item Let $\kappa^{(0)}=g(P(\mathcal{K}\vert \textbf{y},\textbf{x},\theta))$ where $g$ is a function chosen in advanced for providing a value from a distribution. Using MCMC, obtain $M$ samples from the posterior distribution $[\beta\vert \mathcal{K}=\kappa^{(0)},\textbf{y},\textbf{x}]$. These samples will be denoted as $\{\beta\}_{j=0}^M$.
	\item Obtain the point estimates $\beta^{(1)}=g(\{\beta\}_{j=0}^M)$ and similarly $\theta^{(1)}$. Using these values, repeat step 1 to evaluate the distribution $P(\mathcal{K}=k\vert \textbf{y},\textbf{x},\theta^{(1)})$ and obtain $\kappa^{(1)}= g(P(\mathcal{K}\vert \textbf{y},\textbf{x},\theta))$.
	\item Iterate steps 2 and 3 until convergence in the posteriors for $\kappa$ and $\beta$.
	\item Return posterior samples obtained after convergence.
\end{enumerate} In the above procedure, each model is fit conditioned on a single value given by the function $g$. In the remainder of the paper $g$ is taken to give the maximum a-posteriori (MAP) estimate, and the value of $g(P(\mathcal{K}\vert \textbf{y},\textbf{x},\theta))$ at convergence will be referred to as $\kappa^ \text{post}$.

\section{Statistical Validation using Climate Models}\label{validation_study}

This section will use climate models to evaluate the statistical properties of each of the two proposed innovations. In particular there are two sets of choices; the choice between using principal component basis functions versus Laplacian basis functions and the choice between the $\ell_\mathcal{N}$ versus $\ell_{\chi^2}$ likelihood function when modeling $\mathcal{K}$. Ultimately our proposed approach is to use Laplacian basis functions with the $\ell_{\chi^2}$ likelihood model, and we will compare this combination to the principal component parameterization with  both the $\ell_{\chi^2}$ likelihood and the $\ell_{\mathcal{N}}$ likelihood. The combination of Laplacian basis functions with the $\ell_{\mathcal{N}}$ likelihood is omitted here as this method generally performs worse than the others. This is due to the fact that the $\ell_\mathcal{N}$ likelihood is more likely to select a larger number of components, which when applied to the Laplacian parameterization leads to the inclusion of components representing small correlation length scales which are ``noisier'' from the perspective of inference.

\subsection{CMIP6 Data}\label{cmip6_data_sec}

To evaluate these methods, climate model simulation data  was used from the 6th version of the Climate Model Inter-comparison product (CMIP6) \citep{eyring2016overview}. To estimate the covariance structure, pre-industrial control (piControl) scenarios were used, and for the forced component historical simulations were used. The former scenarios are run without any greenhouse gas forcing present and as such it can be assumed that they are distributed according to natural variability. The latter scenarios are run with greenhouse gases at historical concentrations with other forcings such as aerosols, ozone, volcanoes, and solar cycles also present. Monthly ``tas'', or near-surface air-temperature, datasets were used. When necessary datasets were re-gridded to the uniform $5^\circ\times 5^\circ$ latitude-longitude grid over the Earth's surface. Control runs were broken into 25-year segments, and for each segment 25-year trend coefficients were computed at each grid-cell using linear regression. For historical runs regression coefficients were computed over the period 1990-2015. 

Denote pre-industrial control and historical climate models as $\mathcal{P}_c$ and $\mathcal{H}_f$ respectively, where $1\leq c\leq N_{\mathcal{P}}$ and $1\leq f\leq N_\mathcal{H}$.  Processing the climate model data yields trend fields $\mathcal{P}_c=\{\textbf{z}_1,\dots,\textbf{z}_{n_{\mathcal{P}_c}}\}$ and $\mathcal{H}_f=\{\textbf{x},\dots,\textbf{x}_{n_{\mathcal{H}_f}}\}$ for each $c$ and $f$. Attention has been restricted to model scenarios where $n_{\mathcal{P}_c}>5$ and $n_{\mathcal{H}_f}>5$ to ensure that there is sufficient data for testing the inference procedure. Under this restriction $N_\mathcal{H}=7$ and $N_\mathcal{P}=16$ as shown in Table \ref{climmod_summary}. Over the $7$ historical models, there is  $\sum_{f=1}^{N_\mathcal{H}}n_{\mathcal{H}_f}=266$ trend fields in total.

\begin{figure} 
	\centering
	\includegraphics[width=\linewidth]{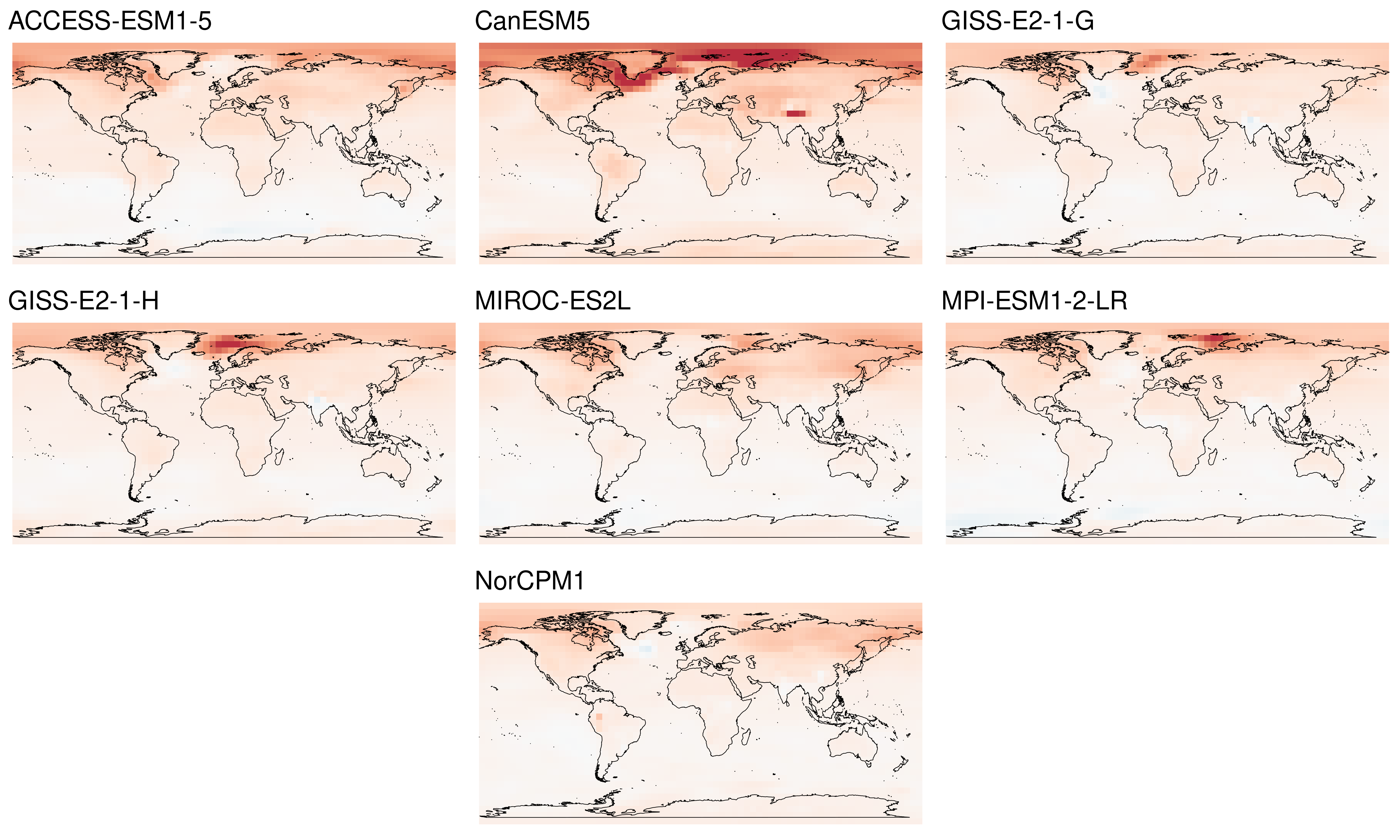}
	
	\includegraphics[width=.8\linewidth]{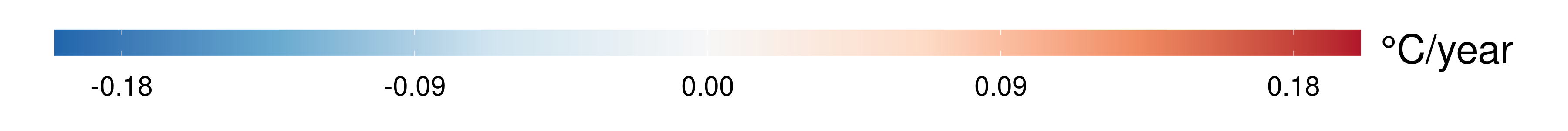} 
	\caption{Mean forced signal for each of the 7 historical climate model ensembles.}\label{histfield}
\end{figure}

Although individual ensemble members will be used in place of the observations in the validation study described in the next section, ensemble means will be used for estimating the forced pattern. For a fixed $\mathcal{H}_f$, let $\overline{\textbf{x}}=\frac{1}{n_{\mathcal{H}_f}}\sum_i x_i$ denote the ensemble mean and let $\overline{\textbf{x}}_{-k}=\frac{1}{n_{\mathcal{H}_f}-1}\sum_{i\neq k}\textbf{x}_i$ denote the mean over the ensemble with the $k$-th member excluded. Figure \ref{histfield} displays the historical trend fields for each of the seven historical simulations. The variation between these estimated forcing patterns implies an additional source of uncertainty arising from error in the climate simulations, which is in addition to the uncertainty induced by estimating the forced signal from limited runs. As historical ensemble means are used, the latter source of uncertainty is not accounted for here, although it would be straightforward to incorporate this portion into the hierarchical Bayesian model in a way similar to KHS17. While simulation error is also not accounted for in the proposed approach, it will be evaluated in the results of this section through intervals over quantities obtained from applying the statistical frameworks to the different historical models. This is similar to ``simulation error'' implied by heterogeneity in the implied covariance structures over $\mathcal{P}_c$ and visualized in Figures \ref{eoffigure} and \ref{lapcoefs}, which will also be evaluated in the following study. 

\subsection{Study Design}\label{studysec}

For a fixed historical climate model $\mathcal{H}_f$, for any $\textbf{x}_k\in \mathcal{H}_f$ the value of the regression parameter $\beta$ in the Equation $\textbf{x}_k=\beta\textbf{x}_{\mathcal{H}_f}+\epsilon$ is known to be equal to one. This ``true value'' setting can be used to evaluate the statistical properties of the Bayesian optimal fingerprinting procedure using historical vectors as a surrogate for the observations.  To ensure that the true value of $\beta$ is one, the climate model generating the surrogate observations will be kept the same as the model used for estimating the forced signal. However, for each historical model, the climate model generating the pre-industrial control vectors will be alternated over the 16 choices in Table \ref{climmod_summary}. This alternating scheme is done in order to investigate the case where the covariance matrix implied by the control vectors differs from the ``true'' covariance generating the observations. The hypothetical ``true'' covariance matrix describing the real-world observations can be thought of as being at least as different from any particular climate model's covariance structure as the covariance structure of any two climate models are from each other. As such, varying the choices of historical and pre-industrial climate models yields insight into the statistical properties when applying the method with a singular pre-industrial control model to real-world observations.

For the study, statistical properties will be evaluated in terms of coverage rates, root-mean-squared errors (RMSE), and continuous ranked probability scores (CRPS), where the CRPS metric evaluates both the accuracy and precision of the predictive distribution \citep{hersbach2000decomposition}. The validation procedure described below takes as input a choice of either principal component or Laplacian basis vectors, a choice of likelihood function for $\mathcal{K}$ out of the choices $\ell_{\chi^2}$ or $\ell_\mathcal{N}$, and choices for $g$ and $M$ as required by the two-fit procedure. The function $g$ will be chosen to give the MAP estimate of the posterior distribution. With these the study proceeds as follows:
\begin{enumerate}
	\item Choose set of pre-industrial control vectors $\mathcal{P}_c$ for $1\leq c\leq N_{\mathcal{P}}$. 
	\item Choose set of historical vectors $\mathcal{H}_f$ for $1\leq f\leq N_{\mathcal{H}}$
	\item For each $1\leq f\leq n_{\mathcal{H}_f}$, let $\textbf{y}\equiv \textbf{x}_k$ represent the surrogate observations and let $\textbf{x}\equiv\overline{\textbf{x}}_{-k}$, where $\overline{\textbf{x}}_{-k}$ is the average of the $n_{\mathcal{H}_f}-1$ ensemble members excluding the $k$-th. The vector $\overline{\textbf{x}}_{-k}$ represents the estimate of the true forced signal $\textbf{x}$.
	\item Given this $(c,f,k)$ pair, use the surrogate observations, estimated forced pattern, and pre-industrial control runs to fit the posterior distribution $[\beta\vert \textbf{x},\textbf{y},\kappa^{ \text{post}}]$ according to the two-fit procedure described in Section \ref{procedure_sec}. Recall that $\kappa^ \text{post}$ is the MAP value of the posterior distribution for $\mathcal{K}$ at convergence.
	\item From the resulting posterior distribution of $\beta$, calculate  $CI_{c,f,k}=\textbf{1}\{1\in [q_{\beta,0.5},q_{\beta,.95}]\}$ where $q_{\beta,0.5}$ and $q_{\beta,.95}$ denote quantiles. This will be equal to 1 if the $90\%$ posterior credible interval for $\beta$ includes the true value of $\beta=1$. Additionally, calculate the posterior mean of $\beta$, which will be denoted as $\beta_{c,f,k}$, and the CRPS value for the posterior distribution of $\beta$ against the true value of $\beta=1$.
	\item After fitting the model for each $1\leq k\leq n_{\mathcal{H}_f}$, calculate the coverage rate $\overline{CI}_{c,f}=\frac{1}{n_{\mathcal{H}_f}}\sum_{k}{CI_{c,f,k}}$. Also calculate the RMSE scores over posterior means RMSE$_{ij}=\sqrt{\frac{1}{n_{\mathcal{H}_f}}\sum_{k}(\beta_{c,f,k}-1)^2}$.
	\item After obtaining $CI_{c,f}$ for each historical model $\mathcal{H}_f$, obtain the distribution of coverage rates $\mathcal{CR}_c=\{CI_{c,f}\}_j$ as well as the analogous distributions over CRPS and RMSE scores. 
\end{enumerate} The above steps are repeated until  $\mathcal{CR}_c$, as well as distributions over RMSE and CRPS values, are obtained for each $1\leq c\leq N_{\mathcal{P}}$.

\subsection{Validation Study Results}\label{results_sec}

The above-described validation study was run with each MCMC fit having $M=2{,}000$ samples and burn-in period of $1{,}000$ iterations. The study was run for three sets of modeling choices; principal components and $\ell_\mathcal{N}$, principal components and $\ell_{\chi^2}$, and Laplacian components and $\ell_{\chi^2}$. In reported results, the principal component basis function approaches will be referred to as empirical orthogonal function (EOF) approaches as they are often referred in the optimal fingerprinting literature. Overall, for each of the three sets of modeling choices, this validation study required $N_{\mathcal{P}}*\sum_j n_{\mathcal{H}_f}=16*266=4256$ two-step model fits over the $(c,f,k)$ tuples. 

\begin{figure} 
	\centering
	\includegraphics[width=\linewidth]{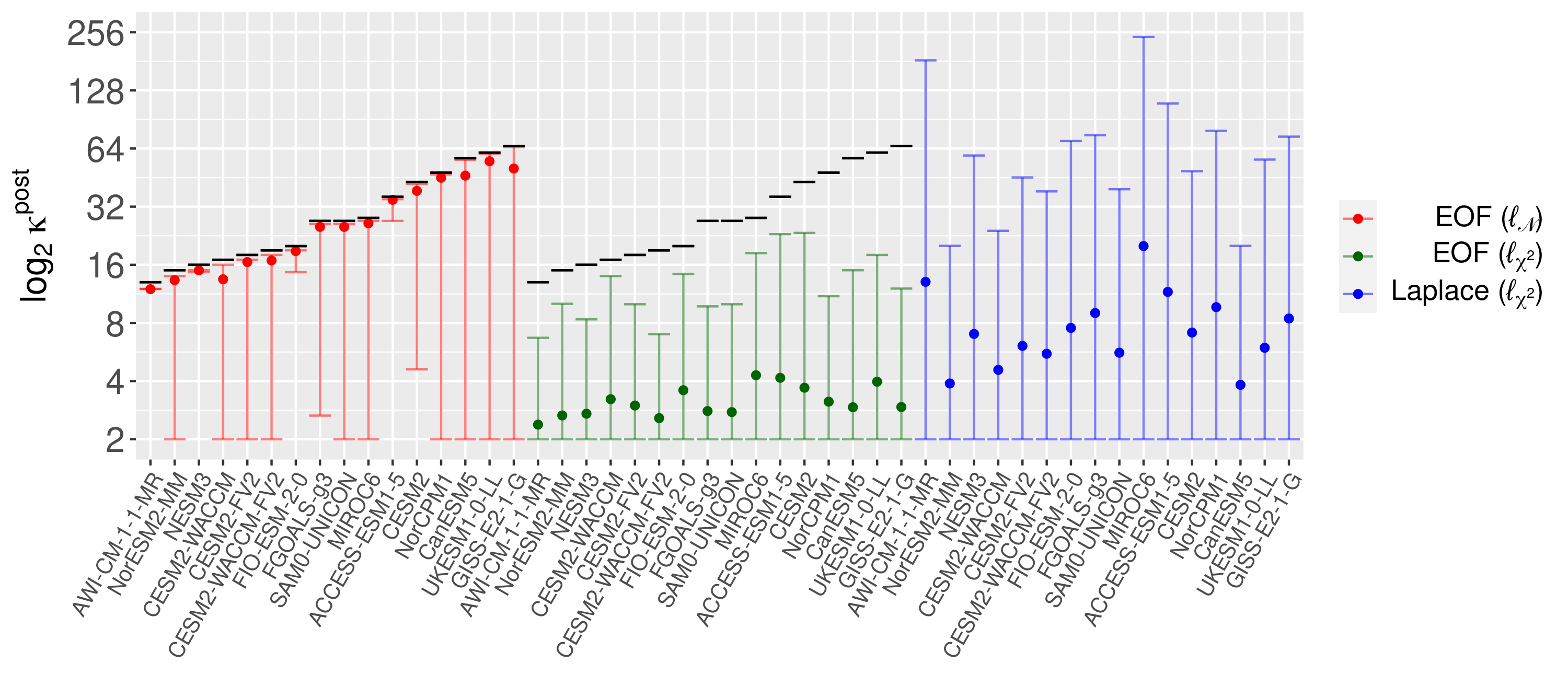}
	\caption{Distribution of $\kappa^{ \text{post}}$ plotted on the $\log_2$-scale (y-axis) over each pre-industrial control model (x-axis) for each of the three statistical frameworks described in Section \ref{studysec}. Means and $90\%$-intervals are shown over the set of all historical ensemble members across the different pre-industrial models.  For the EOF basis function approaches, $\kappa^{ \text{post}}$ is capped at $n_{\mathcal{P}_c}$, or the number of pre-industrial control vectors, which are shown as black lines. Pre-industrial models are ordered according to increasing number of control vectors $n_{\mathcal{P}_c}$.}\label{kappapost}
\end{figure}

Figure \ref{kappapost} displays the distribution of $\kappa^{ \text{post}}$ over the $(f,k)$ tuples for each pre-industrial climate model $1\leq c\leq N_{\mathcal{P}}$. Distributions are visualized through means and $5$-th and $95$-th intervals calculated over the historical ensemble members plotted on the $\log_2$ scale. The models have been ordered on the x-axis by increasing values of $n_{\mathcal{P}_c}$. For the two EOF basis function fits, the number of pre-industrial control runs is shown with black lines, which form a cap on the maximum value of $\kappa^{ \text{post}}$. The maximum potential number of Laplacian basis functions is $n_{ \text{grid}}=2592$, which is beyond the limit of the y-axis.

It can be seen that the EOF-$\ell_{\mathcal{N}}$ version tends to put weight on a larger number of components, and there is a clear trend in increasing $\kappa^{ \text{post}}$ distributions as $n_{\mathcal{P}_c}$ increases. This increasing trend is not the case with the $\ell_{\chi^2}$ versions. It can be seen that in the EOF-$\ell_{\chi^2}$ version, the number of basis functions is generally quite small, with median values often less than eight and always less than the median value under the $\ell_{\mathcal{N}}$ selection method. For each pre-industrial model, the lower-bound on $\kappa^{ \text{post}}$ reaches 2, which is the smallest number of basis functions allowed in the implementation. The Laplace-$\ell_{\chi^2}$ method generally selects more components on average than the EOF-$\ell_{\chi^2}$ method and either more or less than those with the EOF-$\ell_\mathcal{N}$ method, although it should be noted that since these use different sets of basis functions the number chosen is not directly comparable. As an example, when $\kappa^{ \text{post}}=2$,  the EOF method will project on components that differ according to the pre-industrial control model used, the first of which can be seen in Figure \ref{eoffigure}. When $\kappa^{ \text{post}}=2$ with Laplacian basis functions the method will project onto the global mean component and onto the north-south variation component seen in Figure \ref{lapfigure}. In addition, since by construction the EOF approach will maximize the amount of variance explained by each successive component, any number of EOFs will explain an amount of variance that would require a larger number of Laplacians. As such, it is not surprising that a smaller number of EOFs are selected than Laplacians when the adequacy of the covariance estimation is prioritized with the $\ell_{\chi^2}$ parameterization.


\begin{figure} 
	\centering
	\includegraphics[width=\linewidth]{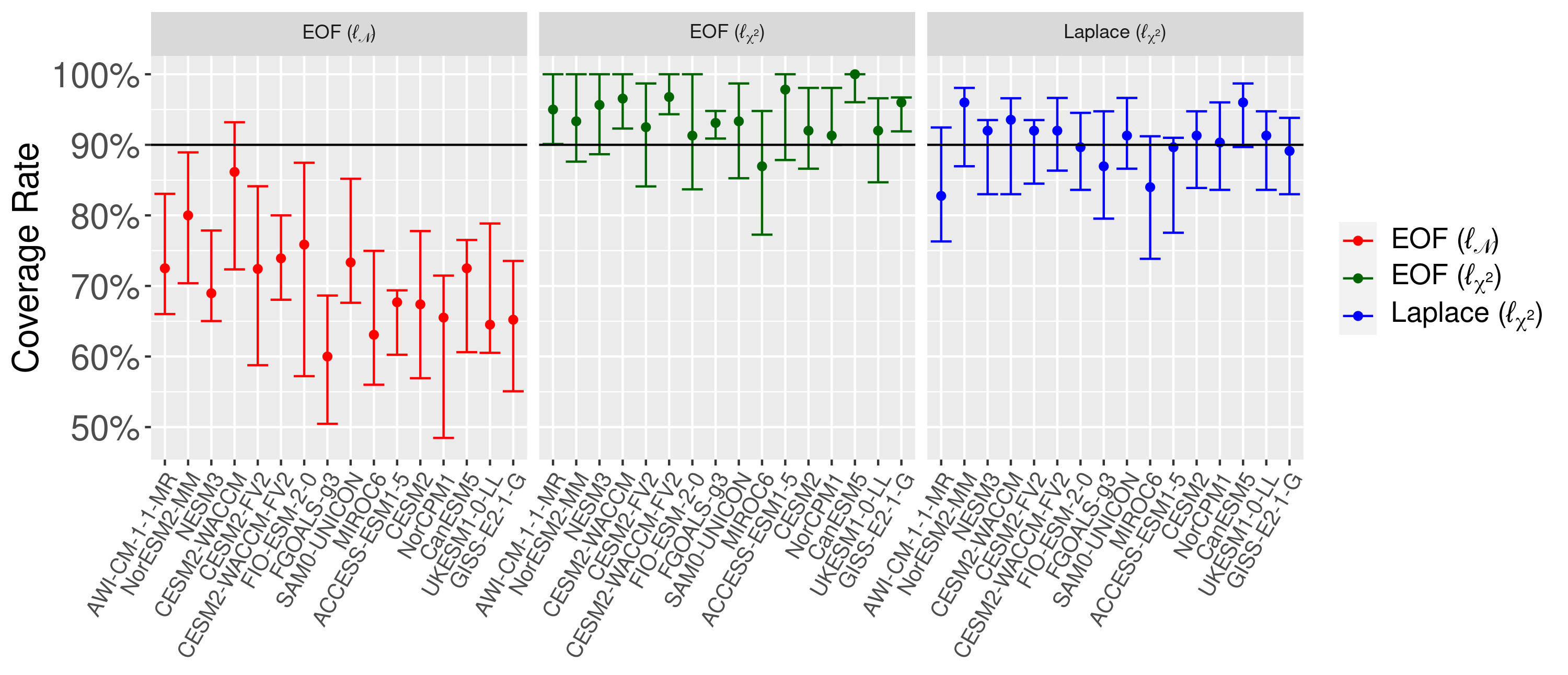}
	\caption{Distribution of $\mathcal{CR}_c$, or the percentage of $90\%$ credible intervals that contain the true value of $\beta=1$, for each pre-industrial control model. Coverage rates were calculated for each historical model separately, and means and $90\%$-intervals are shown over the set of all historical ensemble members. Pre-industrial models on the $x$ axis are ordered according to increasing number of control vectors.}\label{covfig}
\end{figure}

Figure \ref{covfig} shows the distributions of $\mathcal{CR}_{c}$ for each $1\leq c\leq N_{\mathcal{P}}$, recalling that $\mathcal{CR}_c$ is the collection of coverage rates for the $90\%$ Bayesian credible intervals when the ``true'' $\beta$ is known to be one. These are displayed using medians and interquartile ranges over the distribution implied by the historical models, and a horizontal line shows the desired coverage rate of $90\%$. As in Figure \ref{kappapost}, the control models on the x-axis are ordered by increasing number of available control vectors.

For the EOF-$\ell_{\mathcal{N}}$ method, it can be seen that coverage rates are almost always less than 90\% and that under-coverage continues to worsen as $n_{\mathcal{P}_c}$ increases. This property of the EOF-$\ell_{\mathcal{N}}$ method, where the coverage rates decline as the number pre-industrial control models increase, is undesirable as an ideal method would increase in reliability as more data becomes available. From Figure \ref{kappapost} it can be seen that as the number of pre-industrial control vectors increases, this selection approach is more likely to estimate a larger number of components. This leads to inaccurate coverage intervals as the $\ell_{\mathcal{N}}$ likelihood function ``over-fits'' the number of components. This leads to the selection of non-meaningful components that, while improving the fit from the perspective of the likelihood function, leads to over-confidence and sub-optimal accuracy as the number of components gets larger.

Results for the EOF-$\ell_{\chi^2}$ likelihood often achieve accurate coverage but also appear to have a slight pattern of over-coverage. However, the Laplace-$\ell_{\chi^2}$ method achieves well-calibrated coverage rates that do not depend on $n_{\mathcal{P}_c}$, with all of the sixteen intervals including $90\%$. Recall that the $\ell_{\chi^2}$ distributional assumption for $\mathcal{K}$ is intended to prioritize accurate modeling of the covariance components. This choice produces notably better-calibrated coverage rates, with the Laplacian basis parameterization providing a smaller additional improvement over the principal component approach.

\begin{figure} 
	\centering
	\includegraphics[width=\linewidth]{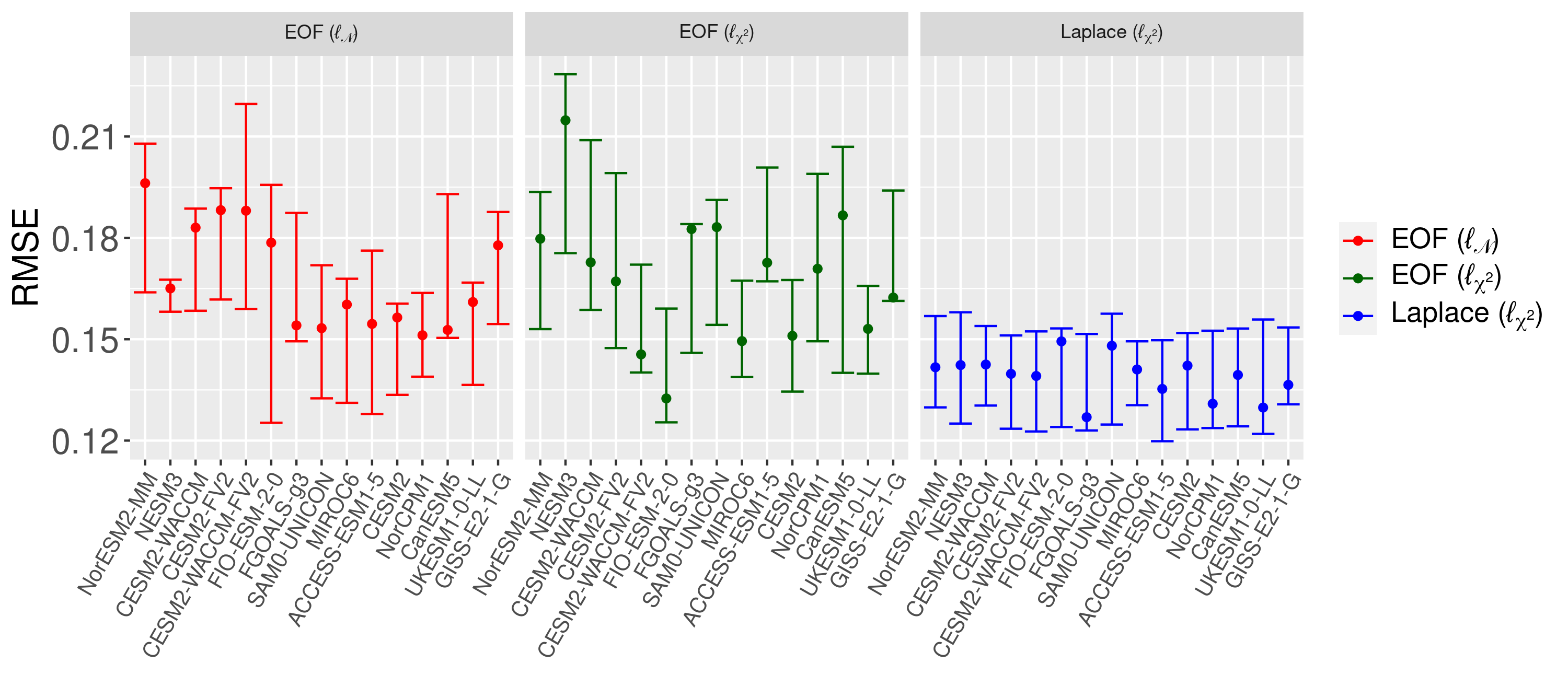}
	\caption{Distribution of RMSE scores calculated using the posterior means of $\beta$ for each ensemble member. Scores are calculated against a ``true'' value of $\beta=1$. Medians and interquartile ranges are shown over the historical climate models for each pre-industrial control model on the x-axis.}\label{rmseplot}
\end{figure}

\begin{figure}[H] 
	\centering
	\includegraphics[width=\linewidth]{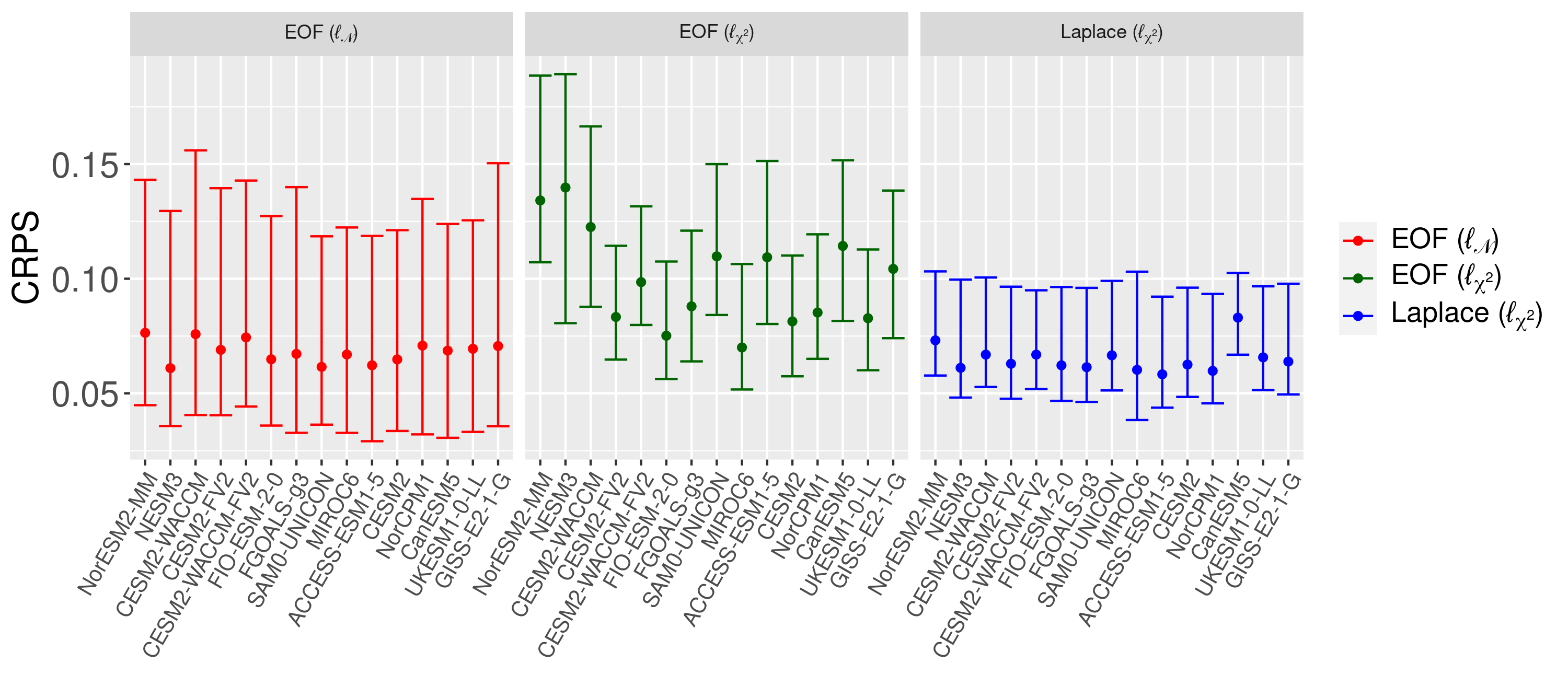}
	\caption{Distribution of CRPS values calculated using the posterior means of $\beta$ for each ensemble member. Scores are calculated against a ``true'' value of $\beta=1$. Medians and interquartile ranges are shown over the historical climate models for each pre-industrial control model on the x-axis.}\label{crpsplot}
\end{figure}

Figure \ref{rmseplot} displays the root mean squared error (RMSE) scores calculated for each historical climate model $\mathcal{H}_f$ using the posterior mean values of $\beta$ from each ensemble member $k$. Medians and interquartile ranges over $\mathcal{H}_f$ are displayed for each pre-industrial climate model $\mathcal{P}_c$ on the x-axis.  It can be seen that, in addition to having well-calibrated coverage rates, the Laplace-$\ell_{\chi^2}$ method generally has the lowest RMSE scores out of the three model selection versions, indicating that this method is able to identify a covariance structure that not only produces accurate recovery of $\beta$ but also produces accurate uncertainty values. The EOF-$\ell_{{\chi^2}}$ has the worst RMSE scores on average and the largest spread both within and between pre-industrial control models. This is unsurprising as this method will generally select a small number of components that are not optimal for estimating $\beta$. The heterogeneity of the first principal components of the covariance structures for each pre-industrial model explains the larger degree of variability in the results for the two EOF methods. The EOF-$\ell_{\mathcal{N}}$ method has a somewhat lower spread and improved scores over the EOF-$\ell_{\chi^2}$ model. This lower spread is due to the larger number of components estimated, so the estimated covariance structures are allowed to have a closer match between the historical model-derived surrogate observations and the control. Both of the EOF methods also exhibit improving scores as $n_{\mathcal{P}_c}$ increases, and while it is not surprising that increasing the amount of data results in improved performance for these methods, this feature is notably absent in the Laplace results, which exhibit low RMSE scores across the pre-industrial control models. 	The Laplacian-$\ell_{\chi^2}$ also has the lowest spread both within and between pre-industrial control models. This is due to the fact that the use of Laplacian basis functions removes both the variability induced by the estimation of principal components from pre-industrial simulation data and the variability induced by the ``mismatch'' between estimated principal components and the unobserved covariance structures implied by the surrogate observations. CRPS values are displayed in Figure \ref{crpsplot}, which shows the same general patterns.

The ``mismatch'' phenomenon can explain the difference in performance between the EOF-$\ell_{\chi^2}$ and the Laplace-$\ell_{\chi^2}$ results. Conceptually the observations can be thought of as being generated from a true covariance matrix that is unknown to the observer. Furthermore, it can be assumed that this theoretical climate model has a variability structure that may be at least as different from any particular climate model as the climate models are from each other. While it has been assumed here that the forced and unforced variability structures are the same, in application and in the validation study this mismatch will occur whenever the observations are from a different real or theoretical covariance structure than the pre-industrial control simulation. This will always occur for the observations due to simulation error and is represented in the validation study through all $(c,f)$ pairs where $\mathcal{P}_c$ and $\mathcal{H}_f$ represent different models. The heterogeneity of the implied covariance structures between the climate models can be visualized for the EOF approach in Figure \ref{eoffigure} and for the Laplace approach in Figure \ref{lapcoefs}. This mismatch entails over-estimation of the variance for lower-number EOFs, since the most significant patterns represented by lower-number EOFs will correspond to less significant and less variable patterns in the observations. This leads to over-coverage when a small number of EOF components are used, as is often the case in the EOF-$\ell_{\chi^2}$ method (Figure \ref{covfig}).  Another way to look at this is that for any two climate models, the first or highest variance EOF of the first model will correspond to a higher-number and lower variance pattern or combination of patterns in the second model. This explains why over-coverage is only present in the lower-$n_{\mathcal{P}_c}$ models in the EOF-$\ell_{\mathcal{N}}$ results, since selecting higher numbers of components allows for patterns to be matched between different EOF structures. Conversely, this mismatch phenomenon forces the EOF-$\ell_{\mathcal{N}}$ method to over-select the number of components, leading to under-coverage as $n_{\mathcal{P}_c}$ grows. 

The use of Laplacian basis functions, on the other hand, does not suffer from mismatching in the structure of the components since they are fixed across historical and pre-industrial models. While under this basis there is still heterogeneity in the component variances as estimated using different climate models, the estimation and uncertainty of these parameters are taken into account in the Bayesian regression framework. This generality and ability to propagate uncertainty in the variance weights yields better-calibrated coverage rates than the two EOF approaches as well as lower RMSE and CRPS values.

A caveat to the use of the Laplacian basis parameterization is that the displayed results only use the $\ell_{\chi^2}$ likelihood for the distribution over $\mathcal{K}$. While not displayed here, when normal likelihood-based component selection methods are used with the Laplacian basis parameterization, the method tends to over-fit the number of components, resulting in under-coverage and worse RMSE scores than the other methods. This issue is not unique to the Laplacian basis parameterization but is due to the larger number of possible components allowed; as can be seen from Figure \ref{covfig} when the $\ell_\mathcal{N}$ method is used with the EOF parameterization, under-coverage becomes more severe as the number of possible components increases. When the effective number of components is increased by adding a diagonal term to the covariance matrix, coverage appears to decrease dramatically. The proposed two-step $\ell_{\chi^2}$ approach provides a solution by modeling the number of components with a likelihood function that prioritizes covariance fit.

In Figures \ref{rmseplot} and \ref{crpsplot}, the results from the pre-industrial climate model AWI-CM-1-1-MR have been omitted since its values in several instances exceeded a reasonable range for the y-axis. The results for this model are reported in Table \ref{awi_results} of the Appendix. For this model the Laplace-$\ell_{\chi^2}$ method has the lowest median RMSE and CRPS values as well as the lowest spread in the values, agreeing with the findings of Figures \ref{rmseplot}  and \ref{crpsplot}

\section{Application to HadCrut Observations}\label{application_sec}

\begin{figure} 
	\centering
	\includegraphics[width=.7\linewidth]{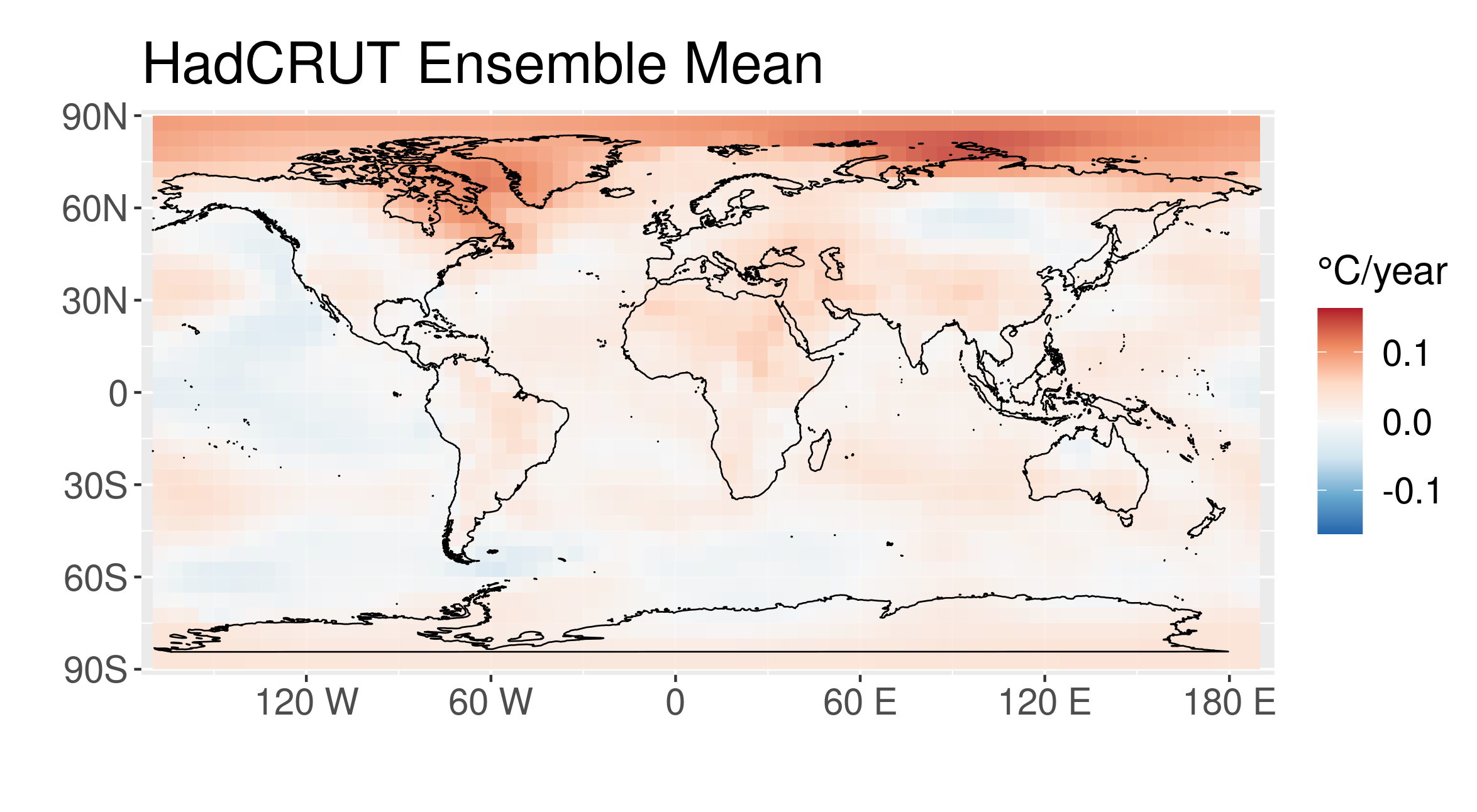} \caption{HadCRUT observational ensemble mean of near-surface air temperature trends over 1990-2015.}\label{hadcrut}
\end{figure}

After evaluating the performance of these three methodological choices within a ``known truth'' context provided by climate model simulations, these frameworks can be applied to real observations to evaluate the substantive effect of the methodological choices. For the observational data, let $\textbf{y}$ be the mean 1990-2015 near-surface air temperature trend field of the 200-member HadCRUT global temperature ensemble  \citep{osborn2021land,morice2021updated}. This field is displayed in Figure \ref{hadcrut}.

\begin{figure}[H]
	\centering
	\includegraphics[width=\linewidth]{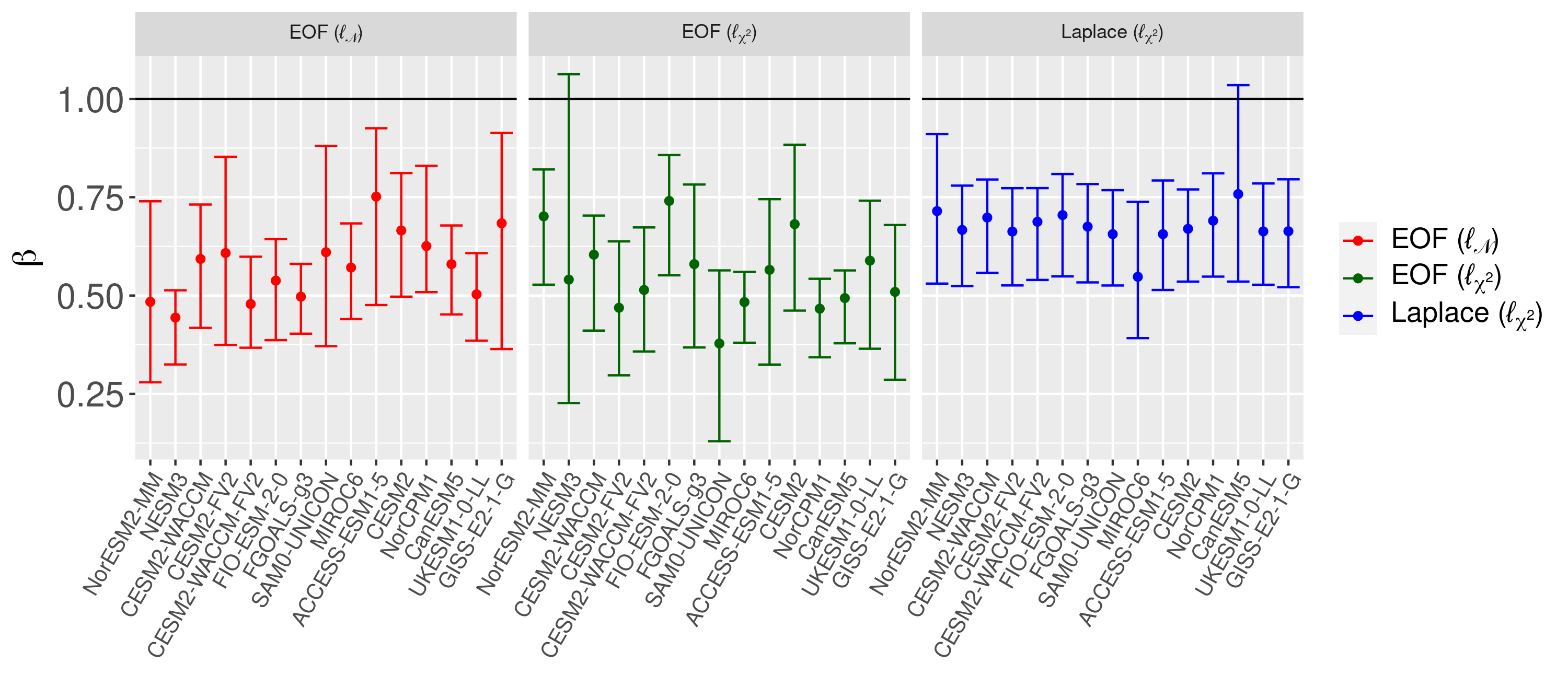}
	\caption{Posterior means of $\beta$ from the application of the Bayesian optimal fingerprinting methods to HadCRUT observational data. Medians and 90\% intervals over the historical model runs are displayed. The black horizontal line is at $\beta=1$.}\label{beta_appplication}
\end{figure}

To quantify the effects of heterogeneity in the choice of climate models, the two-step procedure was applied to the observations with each of the $16*7=112$ pairs of historical and pre-industrial models. Posterior means $\beta_{ \text{post}}$ are displayed in Figure \ref{beta_appplication}. All of the results tend to have a median values notably less than one. This is somewhat surprising, but the consistency of this result over climate model pairs and methods indicates that this feature is inherent in the data. It can be seen that the Laplace-$\ell_{\chi^2}$ method exhibits smaller inter-model variability than the other two approaches. An interesting feature is that the median estimates for the EOF-$\ell_\mathcal{N}$ get larger as $n_{\mathcal{P}_c}$ increases, and appear to get closer to the fairly consistent median values of the Laplacian results. This is a testament to the fact that while the accuracy of the EOF-$\ell_\mathcal{N}$ methods increases with $n_{\mathcal{P}_c}$, the Laplacian method exhibits consistently accurate results regardless of the number of control vectors.

With $\beta_{ \text{post}}$ and posterior standard deviations $\sigma_{\beta; \text{post}}$ the detection problem can be viewed in terms of the ratio $\frac{\beta_{ \text{postmean}}}{\sigma_{\beta; \text{post}}}$, or the distance in posterior standard deviations between the posterior mean of $\beta$ and zero. Greater values of this statistic indicate further evidence for detection. In the hypothesis testing framework, the null hypothesis $H_0:\beta=0$ would be rejected at significance level $\alpha$ if $\frac{\beta}{\sigma_{\beta}}>z_{1-\alpha}$, where $z_{1-\alpha}$ is the $(1-\alpha)$ quantile of the normal distribution. While the displayed results come from a Bayesian framework, similar logic can be employed to make binary detection and attribution conclusions. 

\begin{figure} 
	\centering
	\includegraphics[width=\linewidth]{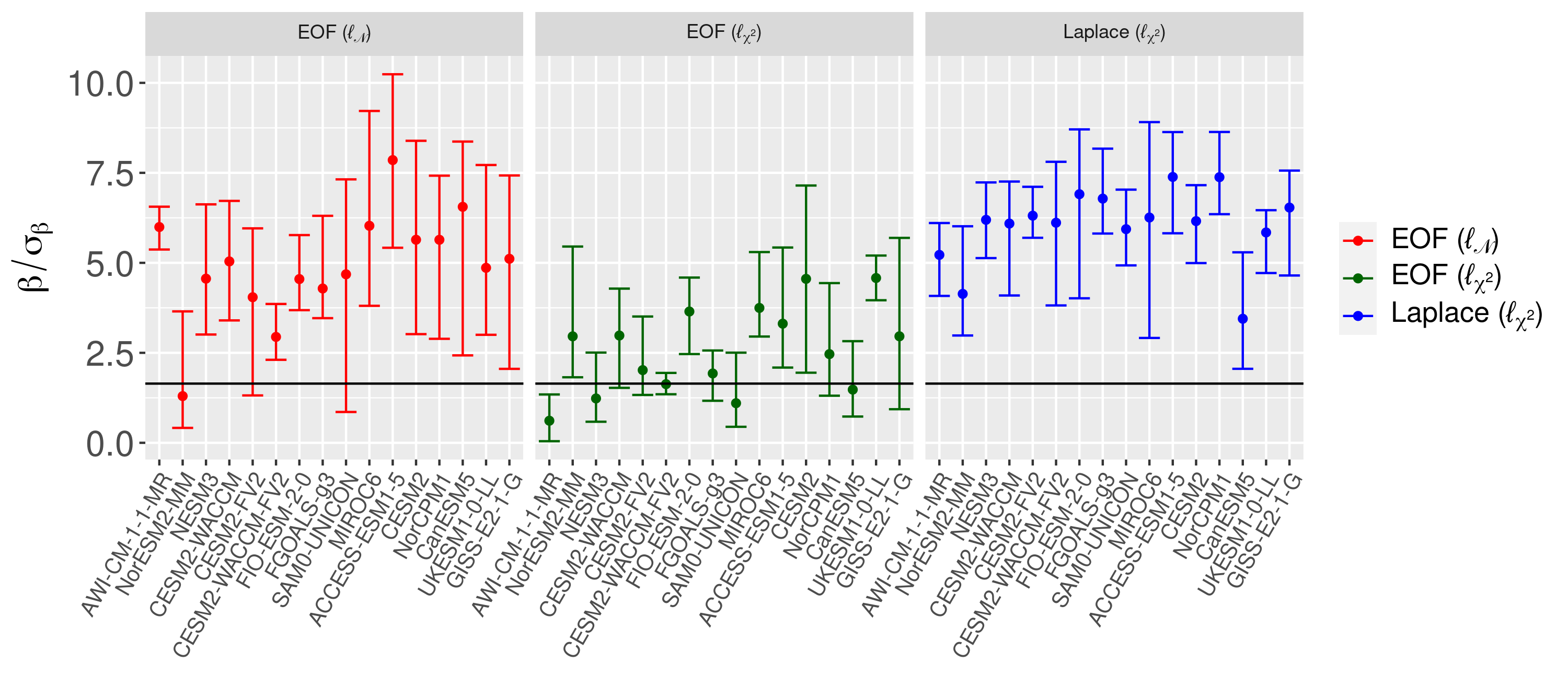}
	\caption{Evidence for ``detection'' of the historical signal in the HadCRUT observational data. The y-axis is presented in terms of number of standard deviations above zero, or $\frac{\beta_{ \text{postmean}}}{\sigma_{\beta; \text{post}}}$. Medians and $90$\% spreads over the historical models are displayed. The black line is at $z_{1-0.05}=1.64$; values above this line would be considered ``detected'' at the 5\% threshold.}\label{detection_results}
\end{figure}

The results for detection are displayed in Figure \ref{detection_results}. Here, medians and $90\%$ intervals are displayed over the set of historical models for each of the pre-industrial control models. The horizontal line at $z_{1-0.05}=1.64$ represents the threshold where the probability of $\beta$ being less than or equal to zero falls below 5\%. It can be seen that the EOF-$\ell_{\chi^2}$ method yields substantially different results than the other two methods in that it frequently fails to find strong evidence for detection. This is due to the fact that this method over-estimates the variance, as evidenced by the validation study of Section \ref{validation_study}. The EOF-$\ell_{\mathcal{N}}$ and Laplace-$\ell_{\chi^2}$ methods both conclude detection over most model pairs, however the Laplacian method exhibits notably less variability between historical and pre-industrial climate models. The higher level of inter-model variability in the EOF approaches is not surprising given the dependence of the EOF representations on the specific control vectors used, and further highlights this advantage of the Laplacian method. 

\begin{figure}[H]
	\centering
	\includegraphics[width=\linewidth]{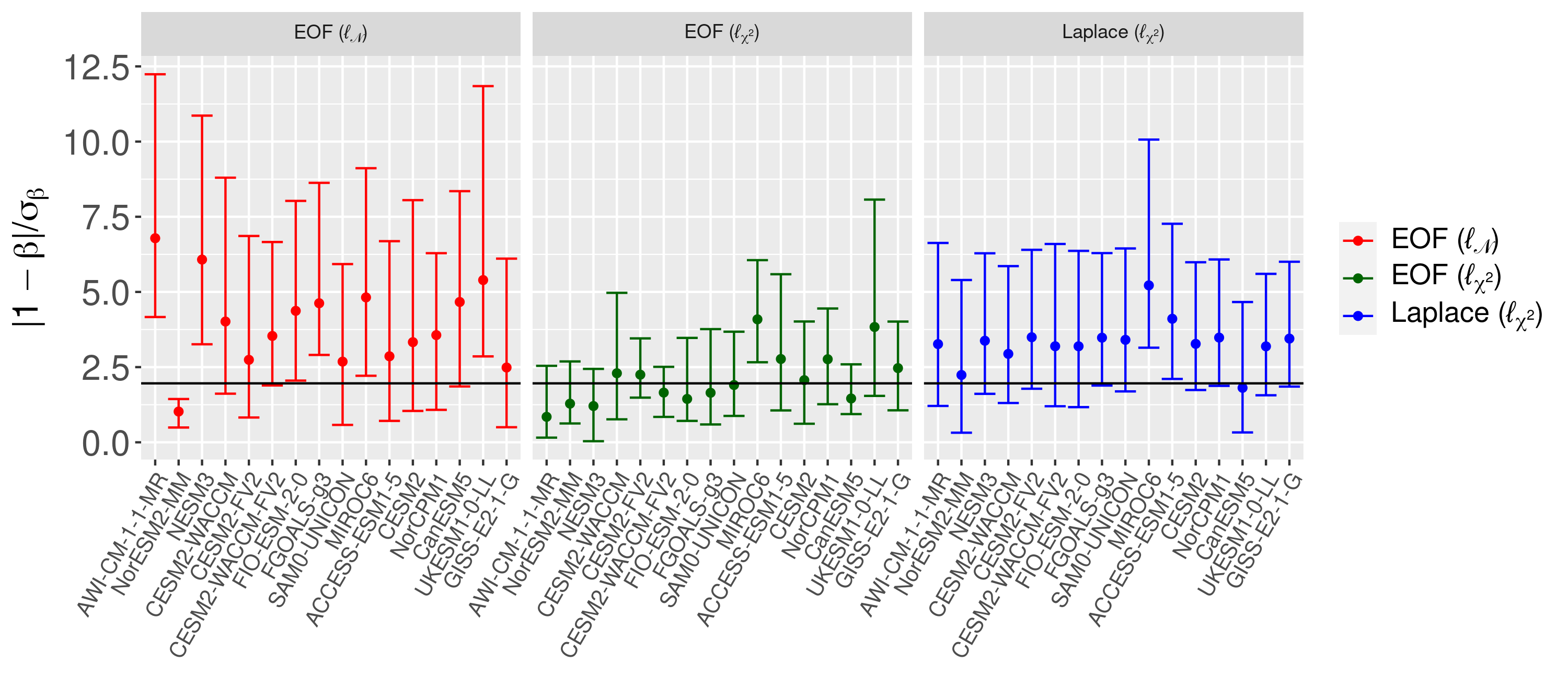}
	\caption{Evidence for the ``attribution'' of trends in the HadCRUT observational data to the historical signal. The y-axis is presented in terms of standard deviations away from one, or $\frac{\vert 1-\beta_{ \text{postmean}}\vert }{\sigma_{\beta; \text{post}}}$. Medians and $90$\% spreads over the historical model results are displayed. The black line is at $z_{0.05/2}=1.96$; values above this line would be considered ``attributed'' at the 5\% threshold.}\label{attribution_results}
\end{figure}

For attribution, the statistic $\frac{\vert 1-\beta_{ \text{post mean}}\vert  }{\sigma_{\beta; \text{post}}}$ indicates the distance in posterior standard deviations between the posterior mean and one. This can be used analogously to the frequentist attribution test, where $\frac{\vert 1-\beta\vert  }{\sigma_{\beta}}$ would be the test statistic for rejecting or failing to reject $H_0:\beta=1$ with a two-sided test. The results are shown in Figure \ref{attribution_results}, where it can be seen that while attribution rates are similar for EOF-$\ell_\mathcal{N}$ and Laplace-$\ell_{\chi^2}$ methods, the inter-model spread is lower for the Laplace-$\ell{\chi^2}$ approach. The EOF-$\ell_{\chi^2}$ approach rarely makes attribution conclusions, which is due to the over-estimated variances obtained with this method.

A numerical summary of the detection and attribution results are displayed in Table \ref{da_results}. Here, ``\# Detected" refers to the percentage of the 112 historical-control combinations whose posterior distribution gives less than a 5\% probability of $\beta$ being less than or equal to zero. This can be seen as analogous to rejecting the null hypothesis of $\beta=0$ at the 5\% level within a frequentist framework. It can be seen here that the Laplace-$\ell_{\chi^2}$ method makes a detection conclusion for every model pair, a testament to this method's consistency and accuracy. The third column shows the percentage of ``attribution'' conclusions, defined as when the $95\%$ credible interval of the posterior distribution contains $\beta=1$. The Laplace-$\ell_{\chi^2}$ method also has the highest percentage of positive attributions, however none of the attribution rates are particularly high due to the fact that $\beta$ is generally estimated to be less than one (see Figure \ref{beta_appplication}). The third column shows the average value of the detection statistic plotted in Figure \ref{detection_results}. The Laplace methods has the highest average confidence in a positive trend, with on average the posterior mean being 6.3 standard deviations greater than zero. This test statistic is slightly higher than that of the EOF-$\ell_\mathcal{N}$ method, and unsurprisingly the EOF-$\ell_{\chi^2}$ method has the lowest value.

\begin{table} 
	\centering
	\begin{tabular}{rllll}
		\hline
		& Method & \# Detected & \# Attributed & Mean $\frac{\overline{\beta}_{ \text{post}}}{\sigma_{\beta; \text{post}}}$ \\ 
		\hline
		& EOF ($\ell_{\mathcal{N}}$) & 100 (89.3\%) & 19 (17\%) & 5.16  \\ 
		& EOF ($\ell_{\chi^2}$) & 70 (62.5\%) & 33 (29.5\%) & 2.13  \\ 
		& Laplace ($\ell_{\chi^2}$) &  112 (100\%) & 34 (30.4\%) & 6.30  \\ 
		\hline
	\end{tabular}
	\caption{Summary of results applied to HadCRUT observational data over each of the 112 control-historical climate model combinations. ``Detection'' and ``attribution'' results were determined using a 5\% posterior probability cutoff. }\label{da_results}
\end{table}

\section{Conclusion}

Our proposed method of using Laplacian basis functions and the $\chi^2$ approach for modeling the number of components presents a further step forward on the path towards more reliable detection and attribution results. The proposed approach has been motivated by two drawbacks in traditional optimal fingerprinting methodology. The first is the use of principal components or EOFs to parameterize the covariance matrix. The fact that the principal components vary meaningfully between different climate models means that their use introduces substantial uncertainty into the optimal fingerprinting procedure which is not taken into account in inference. Our proposed solution is to use Laplacian basis functions to provide a flexible, spatially-coherent parameterization of the covariance matrix whose structure does not need to be estimated from climate simulation data. The second issue is the selection of the number of components, which substantially affects detection and attribution results when using either the principal component or Laplacian covariance parameterizations. The validation study demonstrates that selection approaches based on the normal likelihood parametrization for the number of components produce results that are over-confident, especially when the potential number of components is high. Our proposed solution is to model the number of components with a separate Bayesian model inspired by the $\chi^2$ residual consistency test. This yields inter-dependent models which can be simultaneously fit in an iterative fashion. When the normal likelihood is used for both models, the two-fit procedure can be viewed as equivalent to an integrated Bayesian approach. The ``true data'' climate model study has shown that the Laplacian parameterization using the $\chi^2$ model for the number of components has better-calibrated coverage rates and lower error scores than the principal component parameterization. Finally, we have demonstrated these techniques on observational data, where the Laplace-$\chi^2$ method detects the anthropogenic signal in all of the $112$ climate model combinations with an average confidence of $6.30$ standard deviations.

A notable drawback to this approach is the ad-hoc nature of the two-fit method. The two separate but inter-dependent models are each independently able to quantify uncertainty but are not able to propagate uncertainty through each other. This could be solved through an integrated Bayesian framework for both model selection and parameter estimation. Such an approach is taken in the Bayesian framework of KHS17, however in that case the posterior distribution for the number of principal components is based on the normal likelihood model, which has been shown in Section \ref{results_sec} to give sub-optimal coverage rates. Future work will investigate the potential for a Bayesian framework that can combine the two models proposed here in an integrated fashion.

Another limitation to the proposed approach is that the model selection technique only allows for consecutive sets of components to be included. Such an approach is sensible in the principal component basis parameterization, where successive components have lower empirical variances. In the Laplacian case, however, the relationship of variance to component number is not monotonic as can be seen in Figure \ref{lapcoefs}. Intuitively, when looking at the first nine components in Figure \ref{lapfigure} it could, for example, be sensible to include the first, second, and fifth components (the mean, north-south, and pole-equatorial patterns respectively) while excluding the more longitudinally correlated third and fourth patterns. One potential future approach to address this would be to have a model selection parameter for each component, although this would come with the natural difficulty in that the joint posterior distributions for such parameters would be high-dimensional when the number of potential components is large. Another approach could be to re-order the Laplacians prior to fitting the model, although this would have the undesirable effect of introducing a new source of variability that would not be accounted for in the framework. 

This paper has focused on quantifying uncertainty in the covariance matrix through the form and number of basis functions. To this end other important sources of uncertainty that are accounted for in other optimal fingerprinting approaches have been ignored, including uncertainty in estimating the forced component and observational uncertainty. Both of these sources of estimation uncertainty can be incorporated into the Bayesian model presented here in a straightforward fashion. A more challenging source of uncertainty is climate model error. In the validation study and application, the methods were applied to only a singular pair of historical and pre-industrial control climate models at a time, and the results show a considerable amount of variability across climate model choice. The observations can be viewed as being generated by a covariance structure which is unknown to us but is assumed to be a member of the distribution of covariance structures implied by the spread of inter-model variability. Under this perspective a Bayesian model could be developed which incorporates data from multiple climate models and is able to simultaneously quantify within-model estimation uncertainty and between-model structural uncertainty. The Laplacian covariance matrix parameterization provides a starting point for such a model, as the fact that their form is constant across climate models means that the covariance structures can be pooled within the context of a hierarchical Bayesian framework. This extension has the potential to considerably increase the robustness of the optimal fingerprinting methodology presented here and is planned to be pursued in future work.
	
	\begin{appendices}
		
		\section{Validation Results for AWI-CM-1-1-MR}\label{secA1}
		
\begin{table}[ht]
	\centering
	\begin{tabular}{rlllrrr}
		\hline
		& Model & Method & Metric & $q_{.05}$ & $q_{.50}$ & $q_{.95}$ \\ 
		\hline
		& AWI-CM-1-1-MR & EOF-$\ell_{\mathcal{N}}$ & RMSE & 0.15 & 0.16 & 0.17 \\ 
		& AWI-CM-1-1-MR & EOF-$\ell_{\chi^2}$ & RMSE & 0.45 & 0.59 & 1.05 \\ 
		& AWI-CM-1-1-MR & Laplace-$\ell_{\chi^2}$ & RMSE & 0.15 & 0.15 & 0.17 \\ 
		& AWI-CM-1-1-MR & EOF-$\ell_{\mathcal{N}}$ & CRPS & 0.04 & 0.08 & 0.13 \\ 
		& AWI-CM-1-1-MR & EOF-$\ell_{\chi^2}$ & CRPS & 0.24 & 0.33 & 0.61 \\ 
		& AWI-CM-1-1-MR & Laplace-$\ell_{\chi^2}$ & CRPS & 0.05 & 0.06 & 0.12 \\ 
		\hline
	\end{tabular}
	\caption{Summary of validation study results from the AWI-CM-1-1-MR pre-industrial runs, which were omitted from the figures due to space constraints.}\label{awi_results}
\end{table}

	\section{\textbf{Declarations}}

\noindent{\textbf{Ethical Approval:}} Not applicable

\noindent{\textbf{Competing Interests:}} No competing interests

\noindent{\textbf{Author's contributions:}} SB performed the research and prepared the manuscript, KM advised and reviewed the manuscript

\noindent\textbf{Funding:} SB was supported by the UCLA dissertation year fellowship program. 

\noindent\textbf{Availability of Data and Materials:}  CMIP6 climate model simulation data was obtained from \url{https://esgf-node.llnl.gov/projects/cmip6/} and HadCRUT observational data was obtained from \url{https://crudata.uea.ac.uk/cru/data/temperature/}. Code for performing the analysis has been made available at \url{https://github.com/samjbaugh/DA_BayesLaplace}.
	 	
		
		
		
	\end{appendices}
	

	\bibliography{newbib}
	
	
\end{document}